\documentclass[usenatbib,useAMS]{mn2e}
\usepackage{graphicx}
\usepackage{aas_macros}
\usepackage{amssymb}
\usepackage[pdftex]{}
\usepackage[colorlinks=true]{hyperref}

\hypersetup{
  colorlinks,
  citecolor=blue,
  linkcolor=blue}

\newcommand{\ergs}{erg s$^{-1}$}
\newcommand{\Whz}{W Hz$^{-1}$}
\newcommand{\mics}{$\mu$m}
\newcommand{\msun}{M$_{\odot}$}

\newcommand{\lrad}{L$_{\rm 1.4}$}
\newcommand{\lsix}{L$_{6 \mu m}$}

\title[Red QSOs with e-MERLIN]{Fundamental differences in the radio properties of red and blue quasars: kiloparsec-scale structures
revealed by e-MERLIN}

\author[D. J. Rosario et al.]
{D. J.~Rosario$^{1}$,
D. M.~Alexander$^{1}$,
J. Moldon$^{2,3}$,
L. Klindt$^{1}$,
A. P.~Thomson$^{3}$,
\newauthor 
L. Morabito$^{1}$,
V. A.~Fawcett$^{1}$,
C. M.~Harrison$^{4}$
\\
$^1$Centre for Extragalactic Astronomy, Department of Physics, Durham University, South Road, DH1 3LE, Durham, UK \\
$^2$Instituto de Astrof\'isica de Andaluc\'ia (IAA, CSIC), Glorieta de las Astronom\'ia, s/n, E-18008 Granada, Spain \\
$^3$Jodrell Bank Centre for Astrophysics, School of Physics and Astronomy, University of Manchester, Manchester M13 9PL, UK \\
$^4$School of Mathematics, Statistics and Physics, Newcastle University, Newcastle Upon Tyne, NE1 7RU, United Kindgom \\
}

\begin{document}

\maketitle

\begin{abstract}

Red quasi-stellar objects (QSOs) are a subset of the quasar population with colours consistent with reddening due to intervening dust. 
Recent work has demonstrated that red QSOs show special radio properties that fundamentally distinguish them from 
normal blue QSOs, specifically a higher incidence of low-power
radio emission (1.4 GHz luminosities \lrad $\approx 10^{25}$ -- $10^{27}$ \Whz) that is physically compact when imaged by
arcsecond-resolution radio surveys such as FIRST. In this work, we present e-MERLIN imaging of a set of intermediate-redshift ($1.0<z<1.55$),
luminous (bolometric luminosities L$_{bol} \approx 10^{46}$ -- $10^{47}$ \ergs)
red and normal QSOs carefully selected to have radio properties that span the range over which red QSOs show the most divergence from 
the general population. With an angular resolution $\times25$ better than FIRST, we resolve structures within the host galaxies of
these QSOs ($> 2$ kpc). 
We report a statistically significant difference in the incidence of extended kpc-scale emission in red QSOs. From an analysis
of the radio size distributions of the sample, we find that the excess radio emission in red QSOs can be attributed to structures
that are confined to galaxy scales ($< 10$ kpc), while we confirm previous results that red and normal QSOs have similar incidences of
radio jets and lobes on circumgalactic or larger scales ($> 10$ kpc). Our results indicate that the primary mechanism that generates
the enhanced radio emission in red QSOs is not directly connected with the nuclear engine or accretion disc, but is likely to arise
from extended components such as AGN-driven jets or winds.

\end{abstract}
\begin{keywords}
quasars: individual --
galaxies: jets --
radio continuum: galaxies --
methods: observational --
techniques: interferometric
\end{keywords}

\section{Introduction}

QSOs (Quasi-stellar objects) are the luminous tail of the population of active galactic nuclei (AGNs) and account for a substantial
part of the cosmic mass growth in supermassive black holes (SMBHs). Their distinctive features, high optical luminosities and 
clear spectroscopic signatures of broad emission lines, are the manifestation of an unimpaired line-of-sight view towards the central accretion
engine and its immediate environment. Due to their bright optical emission and relative ease of identification, 
QSOs are valuable signposts of the locations at which the co-evolutionary connections between galaxies and SMBHs are put into place.
Therefore, a thorough understanding of the processes by which QSOs regulate their own growth and influence 
their host galaxies is valuable for our developing picture of the evolution of SMBH scaling relations, and for the 
evolution of massive galaxies as a whole.


The majority of QSOs have blue ultraviolet-to-optical colours consistent with direct thermal emission from an AGN accretion disc peaking in the extreme ultraviolet (effective temperatures $\sim 10^{5}$ K). Large spectroscopic surveys, such as the Sloan Digital Sky Survey (SDSS), have
identified a subset of QSOs which scatter to much redder colours \citep[e.g.][]{richards03}. 
In this work, we will use the abbreviation ``rQSOs'' to refer to red QSOs.

There are a number of possible reasons for
the red colours of rQSOs, such as intrinsically cooler accretion discs, host galaxy light, or doppler-boosted synchrotron emission 
\citep[e.g.][]{benn98, francis01, young08}, but the vast majority of luminous rQSOs are reddened by intervening 
dust \citep{richards03, glikman04, glikman12, rose13, kim18, klindt19}. 

As the colours of rQSOs are not as distinct as those of normal (blue) QSOs, they can be difficult to 
separate photometrically from stars and compact galaxies. In the early days of rQSO research, strong radio emission, a hallmark
of some classes of AGN, was widely used to select and study this sub-population \citep{webster95, ivezic02, white03, glikman04}.
Over time, growing evidence suggested that redder QSOs were associated with a high fraction of radio sources \citep{richards03, white07, georgakakis09, tsai17}, though clear conclusions were elusive due to the complex selection effects that were built into the selection of
rQSOs, including those related to radio emission itself \citep[e.g.][]{richards02}.

In order to make progress in this area, our team has recently undertaken a careful and controlled study of the radio properties of
rQSOs. In a pioneering paper, \citet{klindt19} undertook a controlled study of QSOs selected from the Sloan Digital Sky Survey (SDSS)
and examined their incidence and morphologies using data from the Faint Images of the Radio Sky at Twenty centimetres
survey \citep[FIRST;][]{becker95, helfand15}. We found a much higher detection rate of radio sources among rQSOs compared to equally luminous normal QSOs at the same redshifts. This result is attributable to a substantial excess of compact radio sources among rQSOs, 
with relatively low radio powers (\lrad $ < 10^{27}$ \Whz, where \lrad\ is the integrated rest-frame 1.4 GHz radio luminosity of a radio source). Taking a similar strategy, \citet{rosario20} and \citet{fawcett20} confirmed the results of \citet{klindt19} at regimes of fainter optical and radio luminosities, different radio frequencies, and using higher resolution radio images. 

Our various studies clearly demonstrate that rQSOs have distinct radio properties which cannot be easily explained by
varying reddening over a parent population of normal QSOs.
The chief conclusion is that rQSOs sustain an excess population of compact radio sources, with radio
sizes smaller than the FIRST survey beamsize (5 arcsecond; $\approx 40$ kpc at $z\sim1$), and with intermediate radio-loudness
In this fashion, rQSOs are fundamentally different from normal QSOs, and their properties suggest a deep connection 
between the dust that reddens them and their radio emission mechanisms.

FIRST covers a large area of sky which enables an accurate statistical assessment of the radio-detection statistics for SDSS QSOs.
The angular resolution of FIRST marks a large improvement over most earlier radio surveys, but it offers only 
limited capabilities for the study of distant QSOs. In particular, over the important redshift interval of $1<z<3$ which marks the epochs at which
most SMBH growth occurs, reliance on radio imaging with resolutions of a few arcseconds restricts studies to scales
$\gtrsim 10$ kpc, larger than the stellar sizes of the massive galaxies that host QSOs \citep[e.g.,][]{vdwel14}. 
Since interstellar reddening requires modest columns of gas and dust which are generally only available within galaxies, 
a clearer understanding of the nature of the radio sources in rQSOs requires radio observations with spatial resolutions of a few kpc or lower.

First steps towards a higher resolution study were undertaken by \citet{fawcett20}. Using images from the 1.4 GHz VLA/Stripe82 survey, 
we demonstrated that rQSOs show a slight excess of radio sources with sizes that were close to the resolution limit ($\approx 1.8$''; 
$\approx 15$ kpc at $z\sim1$). 
This cemented the notion that the key to understanding the rQSO phenomenon lay at even smaller scales, well within the realm of 
the QSO hosts themselves.

Here we present e-MERLIN observations of rQSOs and cQSOs with the express aim of understanding 
their morphologies and radio properties on scales of a few kpc. Our study is designed to identify whether the peculiar morphological
properties of rQSOs extends down to galaxy scales much smaller than those probed by FIRST. In Section \ref{data}, after a discussion of the
sample and ancillary data, we describe the new e-MERLIN observations, their reduction, and the processing we have undertaken
to characterise radio structure from the final images. In Section \ref{results}, we lay out our key measurements and results, and
discuss their implications in Section \ref{discussion} for the incidence, origin, and evolution of kpc-scale radio structure in QSOs.
We summarise our conclusions in Section \ref{conclusions}.

Throughout this work, we assume a concordance cosmology \citep{planck16} and implement it in our calculations 
using software built into the AstroPy \texttt{cosmology} module. 

\section{Data and Methods} \label{data}

The QSOs in this study are selected from the SDSS QSO catalogue based on their properties in the SDSS
and FIRST surveys. We describe their selection and general features below, followed by a detailed treatment of
their e-MERLIN observations and analysis.

\subsection{SDSS colour-selected QSOs} 

\subsubsection{Sample selection} \label{sample_description}

We selected our targets for e-MERLIN imaging from the primary sample defined in \citet{klindt19}. Starting from the SDSS DR7
catalogue of QSOs \citep{schneider10}, \citet{klindt19} applied a set of criteria to minimise the complications of selection biases and allow a refined
comparative framework for their investigation into rQSOs. These involved: 1) the use of the \verb+UNIFORM_TARGET+ flag,
2) the exclusion of pure radio-selected QSOs,  3) a required detection in the WISE W1, W2, \& W3 bands (3.5,
4.6 \& 12 \mics\ respectively) to allow the characterisation of the accretion power the AGN, and 4) the restriction to an
upper redshift of $z=2.4$ to ensure that the colour-selection of QSOs was not affected by the Lyman break. Following \citet{klindt19}, we will 
refer to this pre-selected subset of the DR7 QSO catalogue as the `parent sample'.

At the accretion luminosities typical of high-redshift QSOs from the SDSS, the rest-frame mid-infrared (MIR)
is dominated by emission from nuclear dust heated directly by the AGN. 
Since the MIR is negligibly affected by even moderate levels of dust extinction, MIR luminosities serve as a 
good measure of the nuclear power of these QSOs.
From the WISE photometry available, by construction, for the entire parent sample, we estimate the rest-frame
6 \mics\ luminosity (\lsix) through a log-linear extrapolation of the WISE W2 and W3 fluxes, following
the method described in \citet{klindt19} and used by other papers in this series. 

\citet{klindt19} split QSOs from the parent sample into colour-selected categories based on their observed-frame SDSS
$g-i$ colour distributions as a function of redshift. After grouping the parent sample into
redshift-sorted bins of 1000 QSOs, they identified rQSOs as those that lie in the upper 10th percentile of the $g-i$ colour distribution
in a bin. The rQSOs targeted with eMERLIN are a subset of the rQSOs defined by \citet{klindt19}.

For our control sample of `normal' QSOs (henceforth, cQSOs),
we identify the 50\% of QSOs that lie about the median colour in the aforementioned
redshift-sorted bins. This approach differs from \citet{klindt19}, who only use the 10\% median to define their control sample. 
Our main motivation for implementing an expanded colour definition of the control sample was to allow more target choices 
after selection constraints on radio morphology and redshift, luminosity matching requirements, 
and e-MERLIN observability restrictions (see below). We are able to relax the colour-criteria for cQSOs because
\citet{klindt19} clearly demonstrated that the FIRST radio properties of blue QSOs 
do not differ significantly across the full range of colours shown by
this population, implying that both 10\% and 50\% median cQSOs have indistinguishable radio properties. This has been verified
in more recent work \citep{rosario20,fawcett20}, in which a 50\% median control sample, similar to the selection used in this paper, showed differences
in radio properties to rQSOs that are consistent with a 10\% control. 

A final selection requirement is that of radio compactness, since this is the morphological class in which the largest
differences were found between red and normal QSOs.
We identified compact radio sources through visual inspection of images from the FIRST radio survey following the 
methodology laid out in \citet{klindt19}. In practice, such visual determination of morphology is only accurate for sources with enough
fidelity in the FIRST images. Based on our experience with such visual inspection, 
we used a 1.4 GHz flux density cut of 3 mJy and a peak S/N$>15$ as a rule of thumb. Sources with a flux offset 
greater than a factor of 2 between FIRST and the 20 cm NRAO VLA Sky Survey (NVSS) were also excluded; this
criterion identifies systems in which large extended radio lobes are resolved out by the FIRST imaging \citep{klindt19}.
 
Our final pool of candidate targets were chosen from both colour-selected categories (rQSOs and cQSOs) 
in a small redshift interval ($1.0<z<1.55$), display compact FIRST radio morphologies, and lie in a small interval in radio luminosity 
($25.5 < \log L_{\rm 1.4} < 26.5$ W Hz$^{-1}$). The redshift interval was chosen to minimise 
Malmquist biases.
The radio luminosity interval spans the low end of FIRST-detected radio QSOs at our redshifts of interest, 
a regime where the various differences between red and normal QSOs are most pronounced.
As we discuss in Section \ref{mir_radio}, the relative radio-to-MIR luminosities of 
our targets formally place them in the `radio-loud' regime.

From this pool we refined our targets to a final set of 20 rQSOs and 20 cQSOs that were 
matched in redshift and \lsix\ with a tolerance of 0.05 and 0.2 dex respectively. The SDSS identifiers,
redshifts, \lsix\ and other key properties of the targeted QSOs are listed in Table \ref{emerlin_sample}. 
For brevity, in the rest of the text and figures in this papers, we refer to the targets using a shortened form of the 
SDSS identifier: SDSSJ082314.71+560948.9 is shortened to 0823+5609, and so on.

\subsubsection{Sample properties} \label{sample_properties}

\begin{figure}
\centering 
\includegraphics[width=\columnwidth]{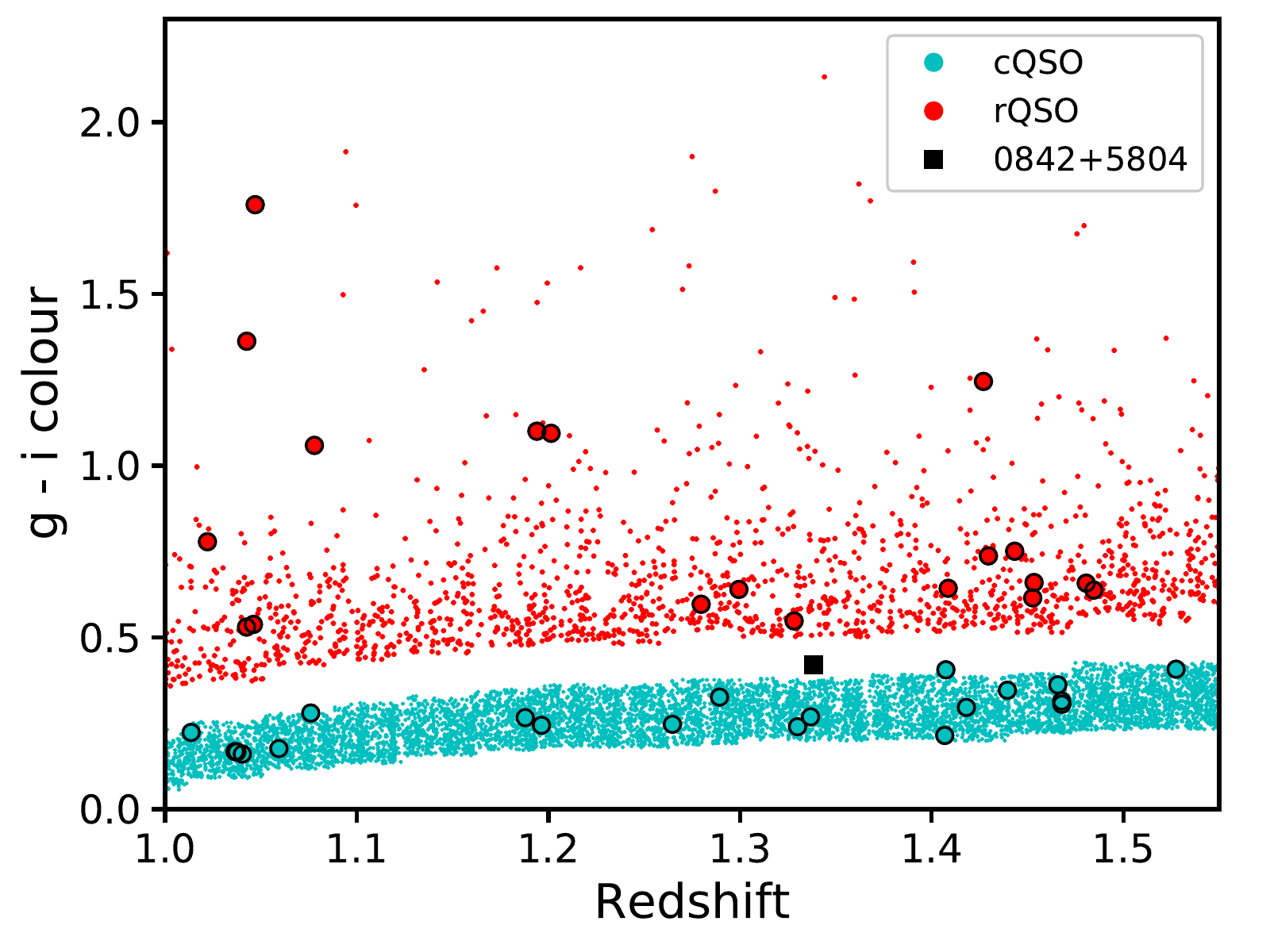}
\caption
{ Observed frame Galactic extinction-corrected $g-i$ colours of SDSS DR7 QSOs in the redshift
range $1.0<z<1.55$ used in this study. The small coloured points are those of the parent sample after applying
the colour selections for rQSOs (red points) and cQSOs (cyan points). More details may be found in Section \ref{sample_description}. 
The targets for this e-MERLIN study as shown using larger points following the same colour scheme as above
except for the single case of 0842+5804 (black square point), which was determined, after correction for Galactic extinction,
to have a $g-i$ colour lying outside our colour-selection windows.
}
\label{gi_redshift}
\end{figure}

In Figure \ref{gi_redshift}, we compare the $g-i$ colours against redshift 
of our e-MERLIN targets to those of SDSS DR7 QSOs from the parent sample of \citet{klindt19}. 
Our targets are identified in the Figure using larger symbols, and the plotting colour scheme of red/cyan for rQSOs/cQSOs
that is used here will the reproduced consistently throughout the rest of the paper. In the post-observation analysis, we discovered
that an incorrect level of Galactic extinction was originally applied to one of our targets (0842+5804), which initially
selected it as an rQSO. With its updated Galactic extinction correction, 
its $g-i$ colour now places it closer to the cQSOs in Figure \ref{gi_redshift}. Hence, we do not consider it as an rQSO for the rest of our analysis.
We discuss this object in more detail in Section \ref{outlier_desc}.

All the rQSOs and 18 of the 20 cQSOs have estimates of black hole masses and Eddington ratios from the 
value-added catalog of spectral properties of SDSS QSOs from \citet{rakshit20}. We find a range in masses ($10^{8}$ --  $10^{10}$ \msun) and 
Eddington ratios ($0.02$ -- $0.83$) in our e-MERLIN targets, consistent with that of the parent sample at their redshifts. 
Since the nature of the radio source emission is broadly related to the Eddington ratios of AGN \citep{best12}, 
we note that our rQSOs and cQSOs have similar median Eddington ratios of 0.12 and 0.15 respectively.
These high values confirm that rQSOs and cQSOs both harbour AGN powered by emission from 
a radiatively-efficient accretion disc. Therefore, differences in their nuclear
engines are unlikely to be the root cause of any differences in their radio properties.

The bright nuclei of our optically-luminous QSO targets prevents a characterisation of their host galaxies. Assuming these
are representative of the hosts of typical $z\sim1$ radio-loud QSOs, we expect them to be mostly bulge-dominated massive
galaxies \citep[e.g.,][]{kukula01}, with moderate-to-strong levels of star-formation 
\citep[$\sim 100$ \msun\ yr$^{-1}$; e.g.,][]{podigachoski15}. This is consistent with the 
extrapolation of trends found at low redshifts, where radiatively-efficient ``high-excitation'' radio AGN are hosted by galaxies
with substantial bulges, and discs that feature on-going star-formation \citep[e.g.,][]{smolcic09, best12}. However, there is no clear evidence
that rQSOs have different host properties from cQSOs at $z\sim1$ \citep{calistro21}, and the contaminating effects of host galaxy light
are not responsible for the red colours of rQSOs \citep[][and discussion below]{klindt19}.

Figure \ref{spectra_comparisons} compares the composite rest-frame UV spectra of the 20 cQSOs and 19 rQSOs. These composites 
are derived by taking a mean of the stack of their SDSS spectra, after correction for Galactic extinction 
by a Milky Way law of \citet{fitzpatrick99}. Before stacking, the 
spectra were normalised to the flux of the QSOs at a rest-frame wavelength of 1.7 \mics, determined from a linear
interpolation of their WISE W1 and W2 fluxes. 
As the rest-frame near infra-red luminosities of rQSOs are minimally affected by the small intrinsic
dust columns that redden these systems, with this choice of spectral normalisation, we are able to 
demonstrate the combined effects of dust extinction as well as reddening.

As mentioned above, the ensemble of rQSOs and cQSOs have approximately equal
MIR luminosities since they are matched in \lsix. Therefore, most of the differences in the overall flux of their median rest-frame UV spectra
is attributable to the differences in intrinsic dust extinction and reddening among the rQSOs. We demonstrate this using the dashed cyan
line in Figure \ref{spectra_comparisons}, which arises from the application of a simple dust screen model to the cQSO composite
adopting an SMC-like extinction law \citep{pei92} set to A$_{V} = 0.2$ mag. Both the normalisation and much of the shape of the rQSO composite is compatible with the effects of mild columns of intervening dust. 

We note differences in some of the spectral features in these composites, such the trough around the C III]$\lambda 1909$ line 
and the equivalent width of the narrow [O II]$\lambda 3727$ line. We refrain from a discussion of these differences in this work; 
a detailed statistical treatment of spectral differences between red and blue QSOs is the topic of a future paper from our team. 

\subsubsection{0842+5804: an outlier} \label{outlier_desc}

\begin{figure}
\centering 
\includegraphics[width=\columnwidth]{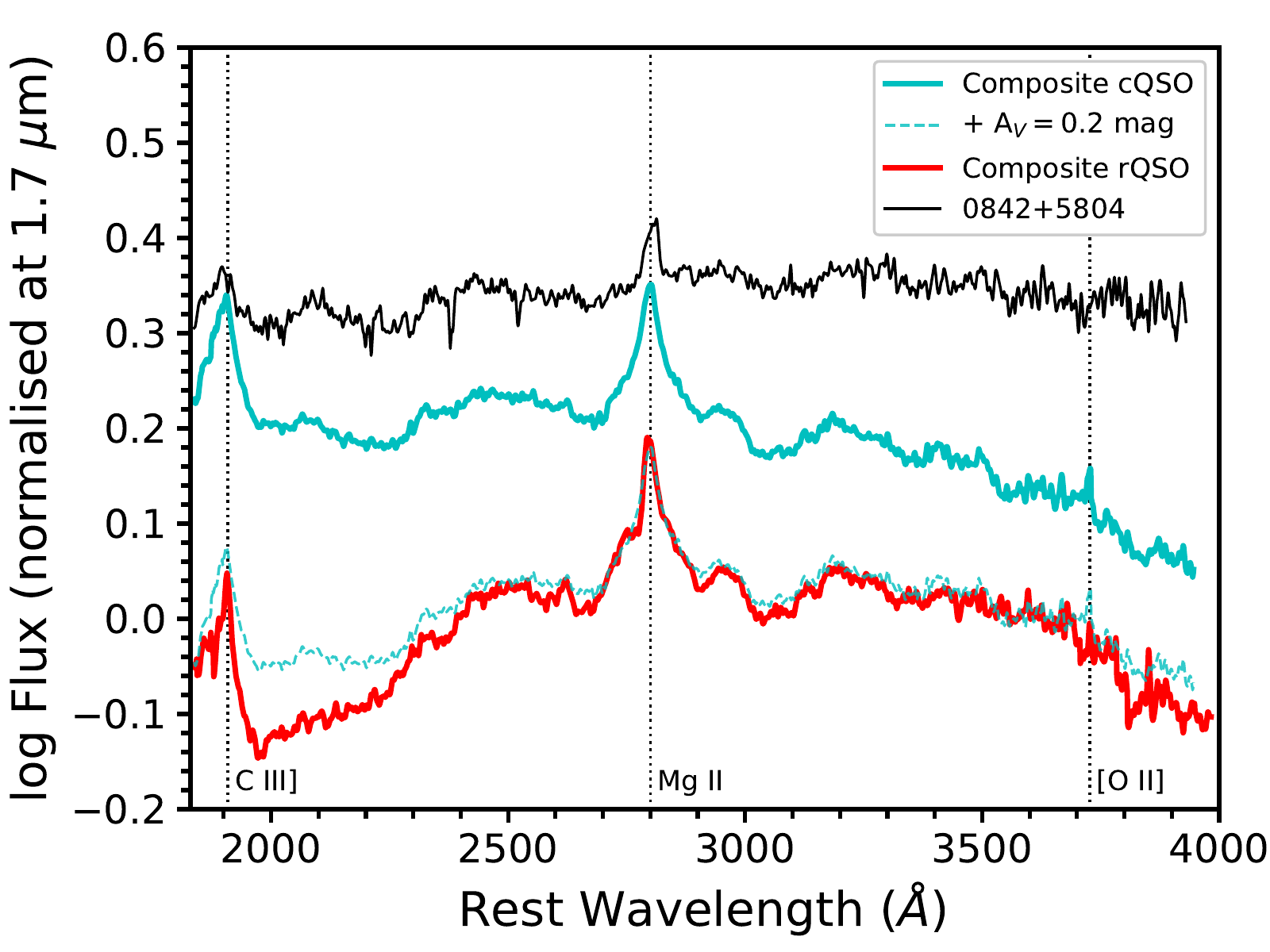}
\caption
{Composite rest-frame UV spectra of rQSOs (red solid) and cQSOs (cyan solid) derived from mean stacks of the SDSS spectra of our targets.
The spectra were corrected for Galactic extinction and normalised to the rest-frame 1.7 \mics\ fluxes of the corresponding 
QSOs before stacking. The cyan dashed line is a version of the cQSO composite that has been reddened by an SMC-like dust
extinction law with A$_{V} = 0.2$ mag.
The spectrum of 0842+5804 is separately shown in black, as it is in outlier in terms of colour and spectral properties.
}
\label{spectra_comparisons}
\end{figure}

0842+5804 was initially selected as an rQSO, but subsequent reanalysis
of its Galactic foreground extinction placed its intrinsic colour closer to those of the cQSOs. Its estimated Galactic visual 
extinction of $A_{V} = 0.275$ is 0.23 mag higher than the median extinction suffered by our targets, and 0.08 mags higher than
the next most extinguished QSO. It lies in the vicinity of a high Galactic latitude dust filament. It is possible that its estimate of extinction,
derived from the average of a 5 arcminute region surrounding the QSO, could be incorrect.

The spectrum of 0842+5804 is plotted in Figure \ref{spectra_comparisons} with a black line to contrast it with the mean composite
spectra of the cQSOs and rQSOs. Its normalisation is higher than both the composites because it has the bluest 
MIR-to-optical colour of our entire sample. Despite this, it shows a UV spectral
shape that is a bit redder than the median cQSOs. It also shows the weakest Mg II$\lambda 2800$ equivalent width
among our targets, as well as weaker than average Fe II band emission. 0842+5804 is an outlier in
both its reddening and its spectral features.

In light of these complexities, we will exclude 0842+5804 from the statistical comparison of rQSOs and cQSOs, which is the
main theme of this work. The exclusion of this one source does not greatly compromise the matched \lsix\ distributions 
that were the basis for the initial sample selection. For completeness, we will continue to include it in the Figures of this Section, 
but we distinguish it using a black plotting colour. It serves as a useful test of the methods we employ to identify extended
structure from our eMERLIN data.

\subsection{Astrometrically accurate QSO positions}

The high angular resolution and milliarcsecond astrometric accuracy
of e-MERLIN data surpasses the astrometric capabilities of the SDSS, which could make a comparison of the optical and radio
positions of the QSOs difficult. We overcame this limitation by searching the GAIA DR2 catalogue \citep{gaiadr2}
for counterparts to our targets. Due to their optical brightness and point-like nature, 39 QSOs have a GAIA 
measurement, providing sub-milliarcsecond positional accuracies. In the various figures of this paper
where the optical positions of the targets are shown, for e.g. Figure \ref{images}, we use these GAIA positions
except for the cQSO 1511+3428, for which only the SDSS position is available. 

\subsection{TGSS 150 MHz fluxes} \label{tgss_radio}

The TIFR GMRT Sky Survey (TGSS) is a 150 MHz imaging survey of 90\% of the sky using the Giant Metrewave Radio Telescope (GMRT),
reaching a median noise level of 3.5 mJy. 
As part of TGSS Alternative Data Release \citep[ADR;][]{intema17}, a catalogue of sources with a detection threshold of $7\sigma$ was
published. As our targets are compact in FIRST, we assume the TGSS source photometry, obtained from images
with a beamsize of $\approx 30$'', captures all of their 150 MHz flux,
though blending with nearby sources may be a source of potential contamination.

39 of our e-MERLIN targets are covered by TGSS; one target, 0828+2731, lies at the edge of a hole in the survey coverage. 
8 are associated with single TGSS detections within 5'', which we take to be their TGSS counterparts.
We use these TGSS fluxes, in addition to default limits for the non-detections, 
in our analyses of radio spectral indices in Section \ref{spectral_indices}.

\begin{table*}
\caption{Key properties of the e-MERLIN targets}
\label{emerlin_sample}
\begin{tabular}{cccccc}
\hline 
\hline 
SDSS identifier & RA (deg) & Dec (deg) & Redshift & \lsix\ & \lrad\ \\
(1) & (2) & (3) & (4) & (5) & (6)  \\
\hline \hline
\multicolumn{6}{l}{Red QSOs} \\ 
\hline 
SDSSJ082314.71+560948.9 & 125.811264 &  56.16362 & 1.4433 & 45.12 & 26.03  \\
SDSSJ082837.75+273136.9 & 127.157333 &  27.52692 & 1.4846 & 45.39 & 25.86  \\
SDSSJ094615.01+254842.0 & 146.562533 &  25.81170 & 1.1940 & 45.78 & 25.91  \\
SDSSJ095114.35+525316.7 & 147.809845 &  52.88796 & 1.4270 & 45.89 & 26.35  \\
SDSSJ100713.68+285348.4 & 151.806977 &  28.89678 & 1.0470 & 46.08 & 25.86  \\
SDSSJ105705.15+311907.8 & 164.271484 &  31.31883 & 1.3282 & 45.74 & 26.33  \\
SDSSJ112220.45+312440.9 & 170.585218 &  31.41139 & 1.4534 & 45.74 & 26.37  \\
SDSSJ114046.80+441609.8 & 175.195022 &  44.26939 & 1.4086 & 45.34 & 26.18  \\
SDSSJ115313.06+565126.3 & 178.304452 &  56.85729 & 1.2014 & 45.84 & 26.49  \\
SDSSJ115924.05+215103.0 & 179.850216 &  21.85083 & 1.0461 & 45.37 & 25.75  \\
SDSSJ120201.91+631759.4 & 180.508038 &  63.29984 & 1.4806 & 45.42 & 25.93  \\
SDSSJ121101.81+222106.7 & 182.757562 &  22.35188 & 1.2994 & 45.99 & 25.79  \\
SDSSJ125146.34+431729.7 & 192.943066 &  43.29158 & 1.4527 & 45.27 & 26.40  \\
SDSSJ131556.37+201701.6 & 198.984945 &  20.28378 & 1.4296 & 45.80 & 25.94  \\
SDSSJ132304.23+394855.0 & 200.767645 &  39.81529 & 1.2797 & 45.48 & 26.12  \\
SDSSJ134236.96+432632.1 & 205.653994 &  43.44224 & 1.0426 & 45.54 & 26.31  \\
SDSSJ141053.19+401618.5 & 212.721690 &  40.27183 & 1.0425 & 44.94 & 25.87  \\
SDSSJ153133.53+452841.6 & 232.889728 &  45.47822 & 1.0221 & 45.46 & 25.69  \\
SDSSJ153555.27+243428.6 & 233.980322 &  24.57461 & 1.0779 & 45.43 & 25.97  \\
\hline 
\multicolumn{6}{l}{Control QSOs} \\
\hline 
SDSSJ074815.44+220059.4 & 117.064346 &  22.01652 & 1.0594 & 46.02 & 25.89  \\
SDSSJ100318.93+272734.3 & 150.828904 &  27.45956 & 1.2893 & 45.85 & 25.95  \\
SDSSJ101935.22+281738.9 & 154.896733 &  28.29416 & 1.0136 & 45.47 & 26.16  \\
SDSSJ103850.89+415512.7 & 159.712031 &  41.92019 & 1.4678 & 45.99 & 25.97  \\
SDSSJ104240.11+483403.4 & 160.667146 &  48.56762 & 1.0375 & 45.54 & 26.15  \\
SDSSJ104620.18+342708.4 & 161.584121 &  34.45234 & 1.1964 & 45.75 & 26.59  \\
SDSSJ105736.17+331545.9 & 164.400726 &  33.26276 & 1.4658 & 45.31 & 25.78  \\
SDSSJ110352.48+584923.5 & 165.968689 &  58.82319 & 1.3300 & 45.63 & 25.95  \\
SDSSJ120335.39+451049.5 & 180.897481 &  45.18043 & 1.0760 & 45.32 & 26.44  \\
SDSSJ122221.37+372335.8 & 185.589086 &  37.39330 & 1.2647 & 45.31 & 26.11  \\
SDSSJ130433.42+320635.5 & 196.139280 &  32.10989 & 1.3368 & 45.68 & 26.17  \\
SDSSJ141027.58+221702.6 & 212.614950 &  22.28406 & 1.4181 & 45.47 & 26.59  \\
SDSSJ142824.76+291606.7 & 217.103202 &  29.26856 & 1.0365 & 45.21 & 26.44  \\
SDSSJ143249.54+292505.7 & 218.206450 &  29.41825 & 1.0403 & 45.38 & 26.10  \\
SDSSJ151100.64+342842.4 & 227.752670$^{a}$ &  34.47845$^{a}$ & 1.4395 & 45.29 & 26.39  \\
SDSSJ153044.08+231013.4 & 232.683666 &  23.17041 & 1.4068 & 46.09 & 25.88  \\
SDSSJ155436.68+285942.5 & 238.652850 &  28.99514 & 1.1879 & 46.01 & 25.84  \\
SDSSJ160245.92+453050.3 & 240.691330 &  45.51398 & 1.4074 & 45.58 & 25.93  \\
SDSSJ163023.12+384700.7 & 247.596378 &  38.78352 & 1.5275 & 45.16 & 26.31  \\
SDSSJ165724.82+204559.5 & 254.353437 &  20.76655 & 1.4679 & 45.48 & 26.20  \\
\hline 
\multicolumn{6}{l}{Outlier} \\ 
\hline 
SDSSJ084255.56+580425.6 & 130.731537 &  58.07379 & 1.3384 & 45.72 & 26.29  \\
\hline 
\hline
\multicolumn{6}{l}{\footnotesize{1, 4: Identifier and redshift from SDSS DR7 QSO catalogue \citep{schneider10}.}} \\
\multicolumn{6}{l}{\footnotesize{2, 3: Coordinates from GAIA, except for \textsuperscript{a}, for which coordinates come from SDSS.}} \\
\multicolumn{6}{l}{\footnotesize{5: Rest-frame 6 \mics\ luminosity from WISE photometry, in log \ergs.}} \\
\multicolumn{6}{l}{\footnotesize{6: Rest-frame 1.4 GHz luminosity from FIRST photometry, in log \Whz.}} \\
\end{tabular}
\end{table*}

\subsection{1.4 GHz e-MERLIN imaging} \label{emerlin_data}

\subsubsection{Observations and reduction} \label{obsred}

\begin{table*}
\caption{Details of the e-MERLIN observations}
\label{emerlin_obs}
\begin{tabular}{ccccc}
\hline 
\hline 
SDSS identifier & Run ID & RMS & Peak  & Extended?\\
(1) & (2) & (3) & (4) & (5)  \\
\hline \hline
\multicolumn{5}{l}{Red QSOs} \\ 
\hline 
SDSSJ082314.71+560948.9 & \verb+CY7220_L_004_20181218+ &   0.07 &  10.55 & No  \\
SDSSJ082837.75+273136.9 & \verb+CY7220_L_002_20181214+ &   0.06 &   1.19 & No  \\
SDSSJ094615.01+254842.0 & \verb+CY7220_L_002_20181214+ &   0.08 &   1.06 & Yes  \\
SDSSJ095114.35+525316.7 & \verb+CY7220_L_004_20181218+ &   0.08 &  10.71 & No  \\
SDSSJ100713.68+285348.4 & \verb+CY7220_L_002_20181214+ &   0.08 &   4.23 & Yes  \\
SDSSJ105705.15+311907.8 & \verb+CY7220_L_001_20181214+ &   0.06 &  13.38 & No  \\
SDSSJ112220.45+312440.9 & \verb+CY7220_L_001_20181214+ &   0.06 &   9.16 & Yes  \\
SDSSJ114046.80+441609.8 & \verb+CY7220_L_004_20181218+ &   0.09 &   5.63 & No  \\
SDSSJ115313.06+565126.3 & \verb+CY7220_L_004_20181218+ &   0.09 &   6.01 & Yes  \\
SDSSJ115924.05+215103.0 & \verb+CY7220_L_001_20181214+ &   0.08 &   9.31 & No  \\
SDSSJ120201.91+631759.4 & \verb+CY7220_L_006_20181220+ &   0.10 &   2.26 & No  \\
SDSSJ121101.81+222106.7 & \verb+CY7220_L_001_20181214+ &   0.06 &   4.82 & No  \\
SDSSJ125146.34+431729.7 & \verb+CY7220_L_006_20181220+ &   0.09 &  34.58 & No  \\
SDSSJ131556.37+201701.6 & \verb+CY7220_L_003_20181216+ &   0.10 &   3.82 & No  \\
SDSSJ132304.23+394855.0 & \verb+CY7220_L_006_20181220+ &   0.06 &  10.84 & No  \\
SDSSJ134236.96+432632.1 & \verb+CY7220_L_006_20181220+ &   0.06 &  20.14 & No  \\
SDSSJ141053.19+401618.5 & \verb+CY7220_L_006_20181220+ &   0.06 &   6.03 & No  \\
SDSSJ153133.53+452841.6 & \verb+CY7220_L_005_20181220+ &   0.07 &   6.26 & No  \\
SDSSJ153555.27+243428.6 & \verb+CY7220_L_005_20181220+ &   0.14 &   4.78 & Yes  \\

\hline 
\multicolumn{5}{l}{Control QSOs} \\
\hline 
SDSSJ074815.44+220059.4 & \verb+CY7220_L_002_20181214+&   0.11 &   4.46 & No  \\
SDSSJ100318.93+272734.3 & \verb+CY7220_L_002_20181214+&   0.08 &   6.31 & No  \\
SDSSJ101935.22+281738.9 & \verb+CY7220_L_002_20181214+&   0.08 &  11.05 & No  \\
SDSSJ103850.89+415512.7 & \verb+CY7220_L_002_20181214+&   0.08 &   5.66 & No  \\
SDSSJ104240.11+483403.4 & \verb+CY7220_L_004_20181218+&   0.09 &  12.20 & No  \\
SDSSJ104620.18+342708.4 & \verb+CY7220_L_001_20181214+&   0.28 &  28.04 & No  \\
SDSSJ105736.17+331545.9 & \verb+CY7220_L_001_20181214+&   0.06 &   2.10 & No  \\
SDSSJ110352.48+584923.5 & \verb+CY7220_L_004_20181218+&   0.12 &   9.15 & No  \\
SDSSJ120335.39+451049.5 & \verb+CY7220_L_006_20181220+&   0.54 &  42.27 & No  \\
SDSSJ122221.37+372335.8 & \verb+CY7220_L_006_20181220+&   0.06 &  10.82 & No  \\
SDSSJ130433.42+320635.5 & \verb+CY7220_L_001_20181214+&   0.09 &   6.81 & No  \\
SDSSJ141027.58+221702.6 & \verb+CY7220_L_003_20181216+&   0.09 &  15.68 & Yes  \\
SDSSJ142824.76+291606.7 & \verb+CY7220_L_007_20190327+&   0.08 &  13.26 & No  \\
SDSSJ143249.54+292505.7 & \verb+CY7220_L_003_20181216+&   0.05 &  17.76 & No  \\
SDSSJ151100.64+342842.4 & \verb+CY7220_L_003_20181216+&   0.06 &   4.28 & Yes  \\
SDSSJ153044.08+231013.4 & \verb+CY7220_L_005_20181220+&   0.14 &   8.73 & No  \\
SDSSJ155436.68+285942.5 & \verb+CY7220_L_003_20181216+&   0.09 &   5.39 & No  \\
SDSSJ160245.92+453050.3 & \verb+CY7220_L_005_20181220+&   0.15 &   7.27 & No  \\
SDSSJ163023.12+384700.7 & \verb+CY7220_L_005_20181220+&   0.07 &   9.90 & No  \\
SDSSJ165724.82+204559.5 & \verb+CY7220_L_005_20181220+&   0.09 &   9.36 & No  \\
\hline 
\multicolumn{5}{l}{Outlier} \\ 
\hline 
SDSSJ084255.56+580425.6 & \verb+CY7220_L_004_20181218+&   0.10 &   6.64 & Yes  \\
\hline 
\hline
\multicolumn{5}{l}{\footnotesize{1: Identifier from SDSS DR7 QSO catalogue \citep{schneider10}.}} \\
\multicolumn{5}{l}{\footnotesize{2: e-MERLIN identifier for the observing run.}} \\
\multicolumn{5}{l}{\footnotesize{3: Root Mean Square (RMS) flux density of the final cleaned image ($3\sigma$ clipped; mJy beam$^{-1}$).}} \\
\multicolumn{5}{l}{\footnotesize{4: Peak flux density of the final cleaned image (mJy beam$^{-1}$).}} \\
\multicolumn{5}{l}{\footnotesize{5: As determined from both visual assessment and core/beam analysis of the eMERLIN images.}} \\
\end{tabular}
\end{table*}

Our 40 targets were observed with the e-MERLIN interferometer using its L-band receivers ($1.23$--$1.74$ GHz).
Six antennas were used for all the observations; the large Lovell telescope was not used due to refurbishment work at
the time, but the sensitivity of the reduced array was adequate for the science goals of this work.
The 512 MHz bandwidth covered by the continuum correlator was split into 8 spectral windows of 128 channels each.
Shortest integration intervals of $3$--$4$ seconds were used.

The observations were carried out over seven separate runs: six in December 2018 and an additional short run
in March 2019 to reobserve a target with better calibration.
Each run consisted of a sequence of calibration scans interspersed with science scans on targets.
To maximise {\it uv} coverage, the $\approx 1$ hour on-source time on each target was split into several sets of typically
three short scans (a few minutes each) separated by $1$--$2$ hours. 3C286 (1331+3030) was used as a flux reference 
for all runs. 

OQ208 (1407+2827) was used as the primary bandpass calibrator and 3C84 (0319+4130)
was used as a pointing calibrator for most of the runs.
During runs \verb+CY7220_L_002_20181214+ and \verb+CY7220_L_004_20181218+, 
strong winds led to the premature parking of one of the antennae in the array
before it could observe the nominal bandpass calibrator.
Therefore, OQ208 was only used as a pointing calibrator in \verb+CY7220_L_004_20181218+, 
and 3C84 served as the primary bandpass reference. 
Complex gain calibrators observed throughout the run were boostrapped for bandpass corrections to 
recover the data from the associated baselines from the troublesome antenna.
In the case of \verb+CY7220_L_002_20181214+, 3C84 was used to calibrate the bandpass of the
antenna after it was brought back to operation.

The radio data were reduced using the newly developed e-MERLIN pipeline, which employs a backbone
of functions from the Common Astronomy Software Applications (CASA) package to flexibly average, flag, calibrate, and
image data from the e-MERLIN correlator. For a few targets with faint phase calibrators, and for the entire \verb+CY7220_L_005_20181220+
run taken under windy conditions, we adapted the parameters from the pipeline (in particular the
averaging intervals for calibration solutions and associated minimum flagging thresholds) to ensure that most of the
data was unflagged and calibration solutions were suitable.

Since the same bandpass and pointing calibrators were used in all seven runs, we examined their calibrated
flux densities to determine the relative flux calibration of our observations. The largest variations were found for
the pointing calibrator which was often observed at a different part of a run sequence than the flux and bandpass calibrators.
We measured the maximal range of the flux scale of the pointing calibrator to be $\approx 9$\%, and we conservatively adopt
this as the relative flux calibration uncertainty of our observations.

Using CASA, we post-processed the pipeline products for the science targets. In some cases, a small amount of manual flagging
was needed to deal with residual RFI and phase errors. We also performed self-calibration on our targets to 
reduce residual phase and amplitude noise and deal with cases where a nearby bright source was affecting the images of the
target. 

Using WSCLEAN v2.6, we produced final images of our targets with a spatial sampling of 0.03"/pixel and a size of 15.36" ($512\times512$ pixels).
The image size was chosen to fully cover the 5" resolution element of the FIRST survey with enough margin to pick up any structure that may
have been barely resolved in FIRST. The choice of pixel size ensured that the images sampled across the minor axis of the highest resolution dataset with at least 3 pixels.

\begin{figure*}
\centering 
\includegraphics[width=0.8\textwidth]{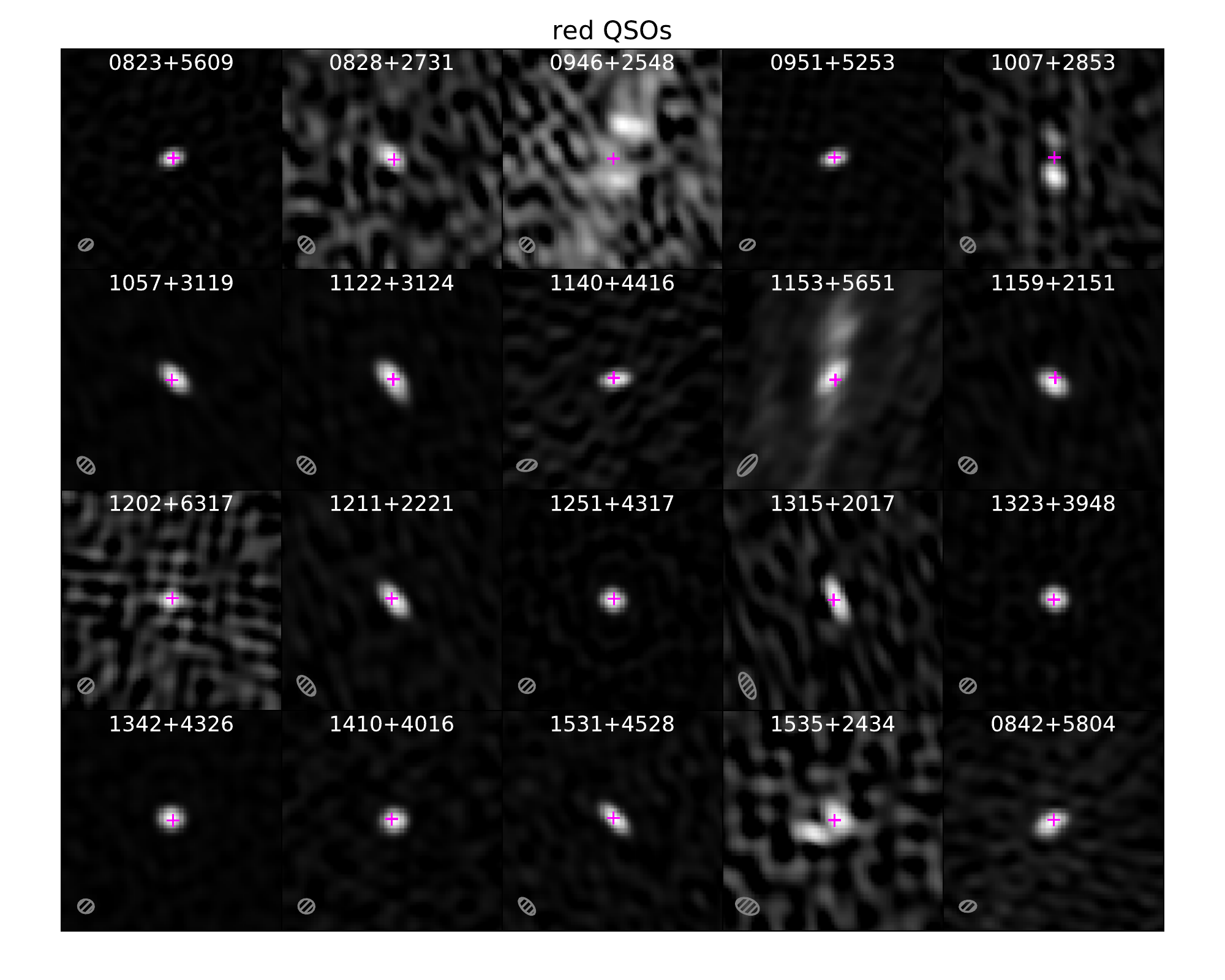}
\includegraphics[width=0.8\textwidth]{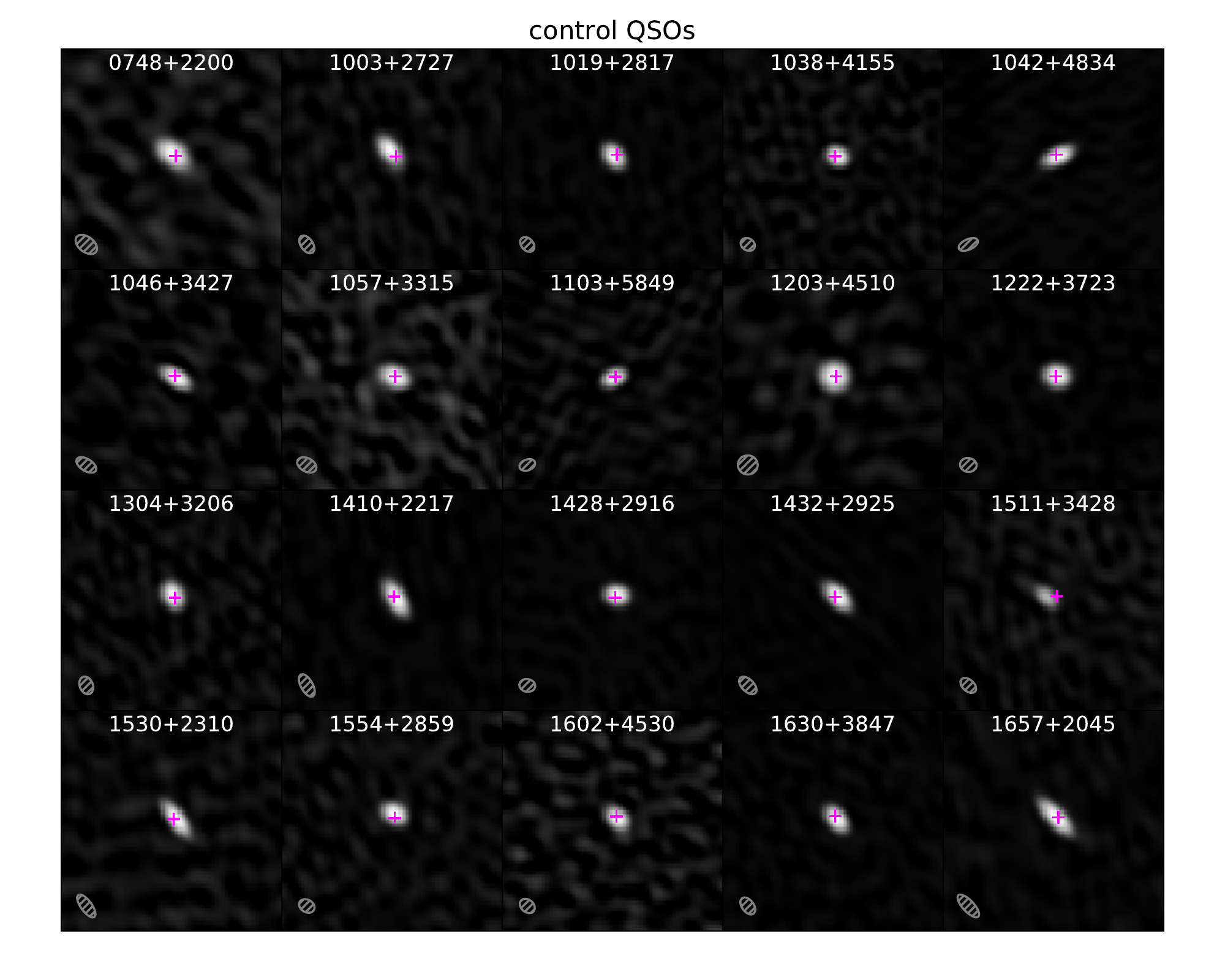}
\caption
{ 2''$\times$2'' stamps from our e-MERLIN images centred on the GAIA-based position of the targeted QSOs (magenta crosses). 
The top panels show the rQSO targets, including the outlier case of 0842+5804 which was initially selected as an rQSO (Section \ref{outlier_desc})
The bottom panels show the cQSO targets. In each panel, a small hatched
grey ellipse shows the size and shape of the restoring beam. 
}
\label{images}
\end{figure*}

In Figure \ref{images}, we show the inner 2''$\times$2'' region of the final e-MERLIN images of the all the QSOs from our sample.
With the long baselines of e-MERLIN, the restoring beams of our images have a median major FWHM of $0\farcs18$, which
translates to projected scales of $1.48$--$1.56$ kpc over the range of redshifts of our targets. Therefore, we resolve radio structures
within the confines of the host galaxies of these QSOs, in contrast to the supergalactic scales ($>40$ kpc)
that are probed by the $5$'' beam of their FIRST images. 

\subsubsection{Gaussian image decomposition} \label{gaussfits}

\begin{table*}
\caption{Results of Gaussian decomposition of the e-MERLIN images}
\begin{tabular}{ccccccc}
\hline 
\hline 
SDSS identifier & RA  & Declination  & Flux density & Major axis & Axis ratio & PA \\
(1) & (2) & (3) & (4) & (5) & (6) & (7) \\
\hline
\multicolumn{7}{c}{Single component fits}\\
\hline
SDSSJ074815.44+220059.4 & 117.064356 &  22.01652 &   4.6 & 0.23 & 1.65 &  49.6 \\
SDSSJ082314.71+560948.9 & 125.811277 &  56.16361 &  11.0 & 0.13 & 1.35 & 110.0 \\
SDSSJ082837.75+273136.9 & 127.157346 &  27.52691 &   1.4 & 0.19 & 1.61 &  50.3 \\
SDSSJ084255.56+580425.6 & 130.731569 &  58.07378 &  11.1 & 0.20 & 1.58 & 120.5 \\
SDSSJ095114.35+525316.7 & 147.809856 &  52.88796 &  11.7 & 0.15 & 1.51 & 107.8 \\
SDSSJ100318.93+272734.3 & 150.828926 &  27.45956 &   7.0 & 0.20 & 1.73 &  34.3 \\
SDSSJ101935.22+281738.9 & 154.896744 &  28.29416 &  11.8 & 0.15 & 1.38 &  38.1 \\
SDSSJ103850.89+415512.7 & 159.712043 &  41.92018 &   5.7 & 0.13 & 1.22 &  67.5 \\
SDSSJ104240.11+483403.4 & 160.667154 &  48.56762 &  13.2 & 0.19 & 1.93 & 115.1 \\
SDSSJ104620.18+342708.4 & 161.584125 &  34.45233 &  25.1 & 0.19 & 1.92 &  58.6 \\
SDSSJ105705.15+311907.8 & 164.271490 &  31.31882 &  14.1 & 0.19 & 1.69 &  45.6 \\
SDSSJ105736.17+331545.9 & 164.400738 &  33.26275 &   2.2 & 0.18 & 1.44 &  69.3 \\
SDSSJ110352.48+584923.5 & 165.968715 &  58.82318 &   8.8 & 0.14 & 1.47 & 113.6 \\
SDSSJ114046.80+441609.8 & 175.195025 &  44.26938 &   5.5 & 0.18 & 1.73 & 100.6 \\
SDSSJ115924.05+215103.0 & 179.850226 &  21.85082 &   9.4 & 0.18 & 1.40 &  53.5 \\
SDSSJ120201.91+631759.4 & 180.508065 &  63.29983 &   2.1 & 0.15 & 1.29 &  88.7 \\
SDSSJ120335.39+451049.5 & 180.897500 &  45.18043 &  40.9 & 0.17 & 1.03 & 105.6 \\
SDSSJ121101.81+222106.7 & 182.757572 &  22.35187 &   5.0 & 0.22 & 1.92 &  38.9 \\
SDSSJ122221.37+372335.8 & 185.589098 &  37.39329 &  12.0 & 0.16 & 1.20 &  78.3 \\
SDSSJ125146.34+431729.7 & 192.943078 &  43.29157 &  33.5 & 0.14 & 1.11 &  72.2 \\
SDSSJ130433.42+320635.5 & 196.139294 &  32.10989 &   6.8 & 0.16 & 1.37 &  22.4 \\
SDSSJ131556.37+201701.6 & 198.984948 &  20.28377 &   3.7 & 0.27 & 2.64 &  22.0 \\
SDSSJ132304.23+394855.0 & 200.767657 &  39.81528 &  11.2 & 0.14 & 1.11 &  68.7 \\
SDSSJ134236.96+432632.1 & 205.654007 &  43.44223 &  21.5 & 0.14 & 1.11 &  85.0 \\
SDSSJ141053.19+401618.5 & 212.721700 &  40.27182 &   6.3 & 0.15 & 1.10 &  91.5 \\
SDSSJ142824.76+291606.7 & 217.103210 &  29.26855 &  13.6 & 0.15 & 1.25 &  79.0 \\
SDSSJ143249.54+292505.7 & 218.206458 &  29.41825 &  18.7 & 0.20 & 1.77 &  44.1 \\
SDSSJ153044.08+231013.4 & 232.683674 &  23.17039 &   8.5 & 0.24 & 2.48 &  35.8 \\
SDSSJ153133.53+452841.6 & 232.889736 &  45.47821 &   7.2 & 0.20 & 2.02 &  43.1 \\
SDSSJ155436.68+285942.5 & 238.652863 &  28.99514 &   6.0 & 0.15 & 1.30 &  58.3 \\
SDSSJ160245.92+453050.3 & 240.691333 &  45.51397 &   7.4 & 0.16 & 1.42 &  45.0 \\
SDSSJ163023.12+384700.7 & 247.596391 &  38.78351 &  10.3 & 0.17 & 1.61 &  38.0 \\
SDSSJ165724.82+204559.5 & 254.353443 &  20.76654 &   9.7 & 0.26 & 2.50 &  43.0 \\
\hline
\hline
\multicolumn{7}{c}{Multiple component fits}\\
\hline
SDSSJ094615.01+254842.0 & 146.562520 &  25.81163 &   0.6 & 0.16 & 1.31 &  65.9 \\
                   & 146.562515 &  25.81177 &   0.9 & 0.15 & 1.49 &  65.2 \\
                   & 146.562468 &  25.81176 &   0.5 & 0.10 & 0.85 & -89.9 \\
                   & 146.562552 &  25.81166 &  13.7 & 2.13 & 5.92 & -19.8 \\
\hline
SDSSJ100713.68+285348.4 & 151.806988 &  28.89673 &   4.9 & 0.17 & 1.41 &  21.8 \\
                   & 151.806990 &  28.89682 &   1.7 & 0.17 & 1.44 &  18.2 \\
\hline
SDSSJ112220.45+312440.9 & 170.585230 &  31.41138 &  10.2 & 0.21 & 1.80 &  41.2 \\
                   & 170.585219 &  31.41135 &   2.0 & 0.28 & 2.60 &  44.7 \\
\hline
SDSSJ115313.06+565126.3 & 178.304467 &  56.85728 &   5.1 & 0.22 & 1.92 & 144.8 \\
                   & 178.304431 &  56.85740 &   1.3 & 0.23 & 2.02 & 120.3 \\
                   & 178.304464 &  56.85729 &  22.3 & 1.90 & 5.17 & -11.8 \\
\hline
SDSSJ141027.58+221702.6 & 212.614961 &  22.28405 &  16.5 & 0.23 & 2.03 &  29.1 \\
                   & 212.615037 &  22.28344 &   3.1 & 0.42 & 2.00 & -75.5 \\
                   & 212.615104 &  22.28371 &   0.3 & 0.11 & 0.79 & -89.9 \\
\hline
SDSSJ151100.64+342842.4 & 227.752709 &  34.47844 &   2.0 & 0.17 & 1.60 &  44.2 \\
                   & 227.754067 &  34.47928 &   7.8 & 0.21 & 1.28 &  44.9 \\
\hline
SDSSJ153555.27+243428.6 & 233.980320 &  24.57460 &   5.9 & 0.23 & 1.50 &  36.4 \\
                   & 233.980388 &  24.57457 &   4.0 & 0.25 & 2.06 &  73.1 \\
\hline
\hline
\multicolumn{7}{l}{\footnotesize{1: Identifier from SDSS DR7 QSO catalogue \citep{schneider10}.}} \\
\multicolumn{7}{l}{\footnotesize{2,3: Peak position of the component on the e-MERLIN image (degrees).}} \\
\multicolumn{7}{l}{\footnotesize{4: Observed e-MERLIN L-band flux density of the component (mJy).}} \\
\multicolumn{7}{l}{\footnotesize{5: Major axis FWHM of the component (arcsec).}}\\
\multicolumn{7}{l}{\footnotesize{6: Major-to-minor axis ratio of the component.}} \\
\multicolumn{7}{l}{\footnotesize{7: Position angle of the component, measured from North ($^{\circ}$).}} \\
\end{tabular}
\label{fit_summary}
\end{table*}

In order to characterise the radio source structure in the e-MERLIN images, we decomposed the images 
into a set of two-dimensional elliptical Gaussian components. For this, we used the AstroPy modelling subpackage, with
minimisation through its implementation of a least-squares Simplex algorithm. We set fitting bounds
on the Gaussian components to ensure positive amplitudes and sizes greater than 50\% of the
corresponding restoring beam\footnote{Fits to the images with the CASA IMFIT 
task gave us very similar results. In this work, we only report our AstroPy-based modelling, as this approach
allowed for greater flexibility in the choice of a robust minimisation algorithm.}. 

In a first pass, we undertook a run of fits over the central 2'' of each image 
adopting only a single Gaussian model which was initialised to match the shape of the restoring beam. 
After examining the fitting residuals visually, we flagged the five cases in which a single Gaussian did not adequately represent
of the source structure. Two more objects showed emission beyond the central 2'' which we also flagged.
These seven images were then subjected to a second round of fits using multiple ($>2$) Gaussian components.

In each of the multiple component fits, we first manually determined the number of components to apply to the fit, and
estimated initial amplitudes, positions, and sizes for each component. After the fits, we examined the residuals
and added or removed components as needed to capture most of the visual source structure. We required between 2 and 4
components to fit all the sources with initial evidence for extended structure. Since the noise in the
e-MERLIN images have a large correlated component, we resorted to a visual judgement of the fit quality rather than
taking a more quantitative metric. This lack of precision is not a major source of uncertainty for this work.

The results of the Gaussian decomposition of the e-MERLIN images are tabulated in Table \ref{fit_summary}, and a visualisation
of the final fits for all the targets is available in the online material (Supplementary Figures 1 \& 2). In Section \ref{visually_extended}, 
we provide a detailed description of the extended radio structure among our targets. 


\subsubsection{Searching for additional extended sources} \label{extension_search}

\begin{figure}
\centering 
\includegraphics[width=\columnwidth]{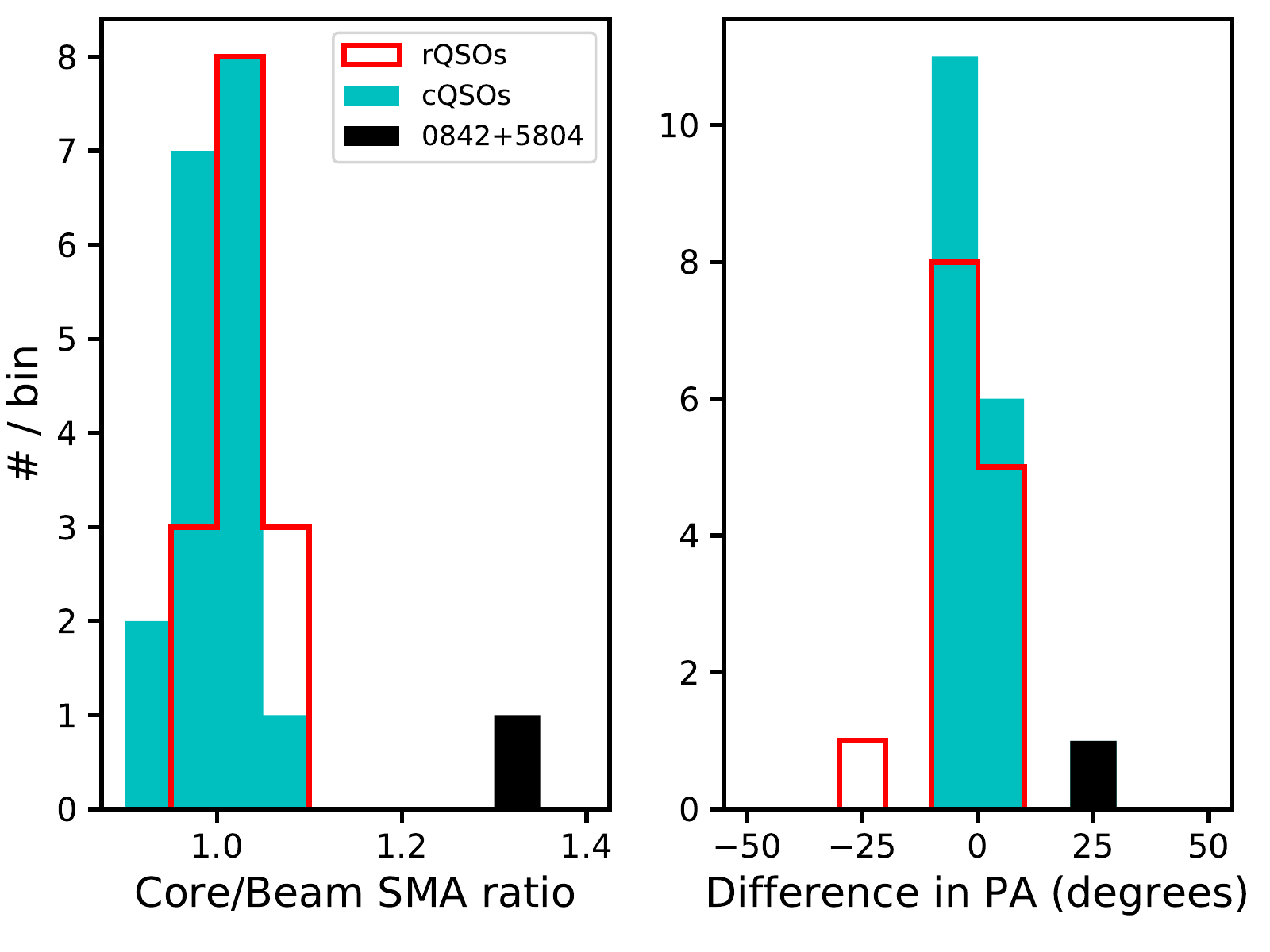}
\caption
{{\bf Left: } The ratio of the semi-major axis of the core component to that of the restoring beam for sources that were visually
determined to be compact in the e-MERLIN images.  Most of the core components have sizes comparable to that of the beam,
except for 0842+5804 (plotted in black). The distribution of core extensions of rQSOs is marginally larger than those of the cQSOs.
{\bf Right: } The difference in position angle between the core components and the restoring beam for sources that were visually
determined to be compact in the e-MERLIN images. Except for 0842+5804 (plotted in black) and a few sources imaged with almost
circular beams, the core components are aligned with their beams. From these two figures, we conclude that the majority of
visually compact cores are unresolved in our e-MERLIN observations.
}
\label{beamcomps}
\end{figure}

To complement the visual analysis described above, we developed an approach to search for additional 
extended source structure at the resolution limit of our images. This was performed only on the 33 sources for which a single component
fit was determined to be an adequate representation of the images. We distinguish between rQSOs and cQSOs in the analysis below,
and also separately include the outlier QSO 0842+5804 because it serves as a useful validation of our technique.

We searched for evidence of extended structure through a comparison of the parameters of the single-component 
Gaussian fits with those of the restoring beam of the images. In the following description, 
we refer to this single component as the `radio core'.

In the left panel of Figure \ref{beamcomps}, we plot the ratios of the semi-major axis of the radio core component 
to that of the beam, splitting rQSOs and cQSOs into different histograms. 
The distributions for most QSOs are centered about unity indicating that these single
component systems are unresolved in our e-MERLIN imaging. 
0842+5804, shown in black in Figure \ref{beamcomps}, is the only visually compact object 
which has a significantly larger core size than the corresponding restoring beam. 
While its radio core is fit well by a single Gaussian, the size of this component is $\approx 30$\% larger than the beam, 
showing that it is definitely extended.

There is a mild suggestion that the ensemble of 
visually compact rQSOs are marginally larger than their beams, by $1.1\pm0.25$\%, in contrast to the ensemble
of cQSOs which are consistent with unity at the same level of uncertainty. However, the distributions of core/beam size
ratios for the two subsets are statistically indistinguishable with our current sample sizes. 

We also compare the best-fit position angles (PAs) of the core components to those of their respective restoring beams
(right panel of Figure \ref{beamcomps}). Most of the systems have core and beam PAs that agree within $\pm10^{\circ}$. One rQSO 
(1202+6317) and one cQSO (1203+4510) have almost circular beams, and their core PAs are extremely uncertain, leading to large
PA differences that are not significant. The only object with a definite difference between the core and beam PA is 
0842+5804. Therefore, the relative core/beam size and PA analyses yield consistent results, 
identifying 0842+5804 as a QSO with extended emission that is not obvious from a visual assessment.  

\begin{figure}
\centering 
\includegraphics[width=\columnwidth]{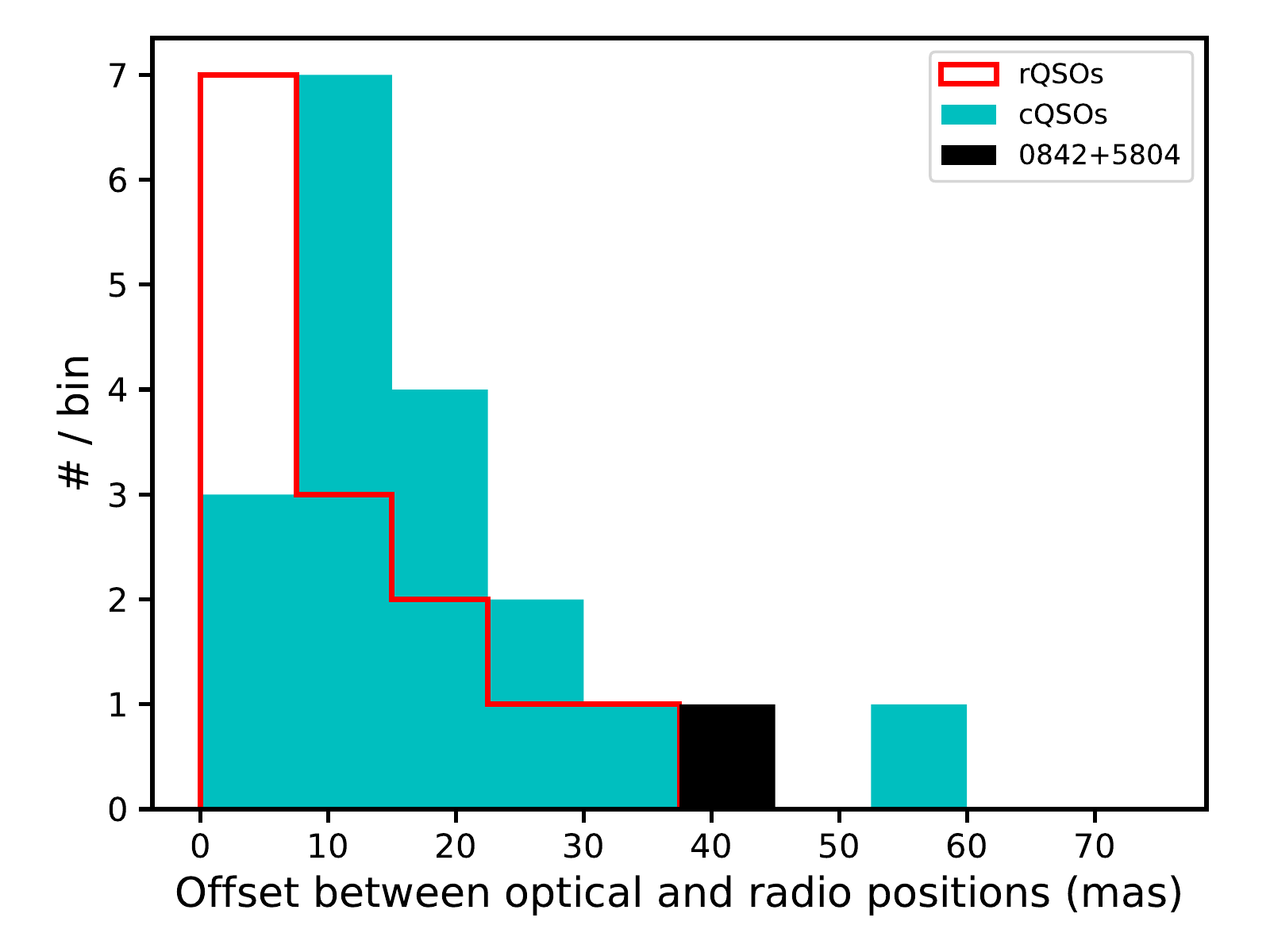}
\caption
{Spatial offsets between the GAIA positions of the optical QSOs and the fitted centres of the 
radio core components for the targets that were visually
determined to be compact in the e-MERLIN images.
}
\label{offsetcomps}
\end{figure}

Taking advantage of the high astrometric accuracy of the e-MERLIN images and the QSO positions from GAIA, 
we search for angular offsets between radio core components and the optical positions of the 
unresolved or single-component sources. As the optical position is centred on the true AGN nucleus,
a systematic measurable difference between the relative optical and radio centroid could indicate that the radio
core contains asymmetric emission below the resolution of the e-MERLIN images, betraying the presence of
radio structure on sub-kpc scales.

We first verified that the optical positional offsets of the radio core components distributed about a zero
mean in both RA and Declination. This ensures that any systematic astrometric error between the e-MERLIN
and GAIA reference frames are negligible, as expected. Translating the optical offsets to an angular distance in milliarcseconds (mas),
we compare, in Figure \ref{offsetcomps}, the distributions of the offsets for our sample, split into colour-selected subsets.
 
We find that the single-component cQSOs and rQSOs show a similar range of offsets between their optical and radio core positions. Most
lie within 30 mas, the size of a single pixel in our images and $\approx 1/6$ of the typical beamwidth of the images. This scatter is consistent
with the errors of the radio core centroids from our Gaussian component fits. There is a mild difference in the shape of the offset
distributions between cQSOs and rQSOs, but we do not consider this to be significant based on a two-sided Kolmogorov-Smirnov (K-S) test.

Two objects show offsets in the long tail of the distribution: 0842+5804 and the cQSO 1003+2727. The evidence for extended structure in
0842+5804 has been presented above, and this may explain its larger offset. 1003+2727, on the other hand, 
has an image that is completely consistent with the restoring beam, yet the respective panel in Figure \ref{images} clearly shows
a visible offset between the centre of its apparently 
unresolved core and the optical position. However, the gain calibration for this particular source was complicated by heavy flagging
of the calibrator scans by the pipeline, and the final astrometric solution may be unreliable. In light of this, we do not place much
weight on the apparent offset, and henceforth treat 1003+2727 as a single component core-dominated source.
Possible higher frequency e-MERLIN observations in the future may help to verify its nature.

\section{Results} \label{results}

\subsection{Visually extended radio sources} \label{visually_extended}

Five of our targets exhibit visually extended emission, both from an examination of the central 
images (Figure \ref{images}) and based on their single-component Gaussian fits: 
0946+2548, 1007+2853, 1122+3124, 1153+5651, and 1535+2434.
All five are rQSOs. In two cQSOs which have unresolved cores, 
additional extended emission is seen on scales larger than the cutouts: 1511+3428 shows a bright radio hotspot 5'' away
from its core, and 1410+2217 has a faint lobe-like structure 2.2" from the core. 

\begin{figure*}
\centering 
\includegraphics[width=\textwidth]{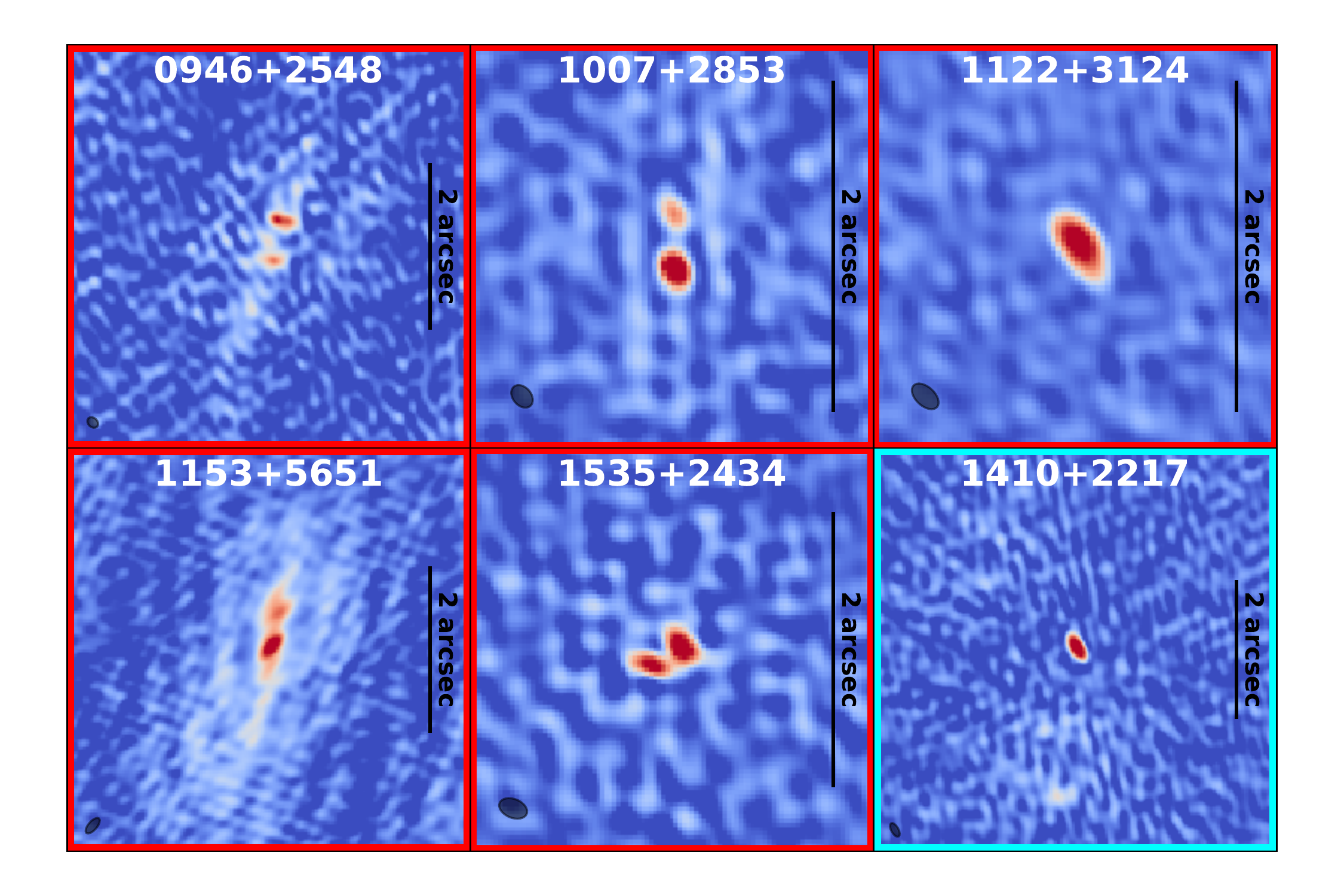}
\caption
{A montage of the visually extended sources from the e-MERLIN observations. A bar corresponding to a scale of 2''
is shown to the right of each panel, which corresponds to the largest angular scale to which the observations are sensitive.
Red or cyan coloured borders indicate whether the source is an rQSO or cQSO, respectively. Section \ref{visually_extended}
discusses these sources in more detail.}
\label{extended_sources}
\end{figure*}

\begin{figure*}
\centering 
\includegraphics[width=\textwidth]{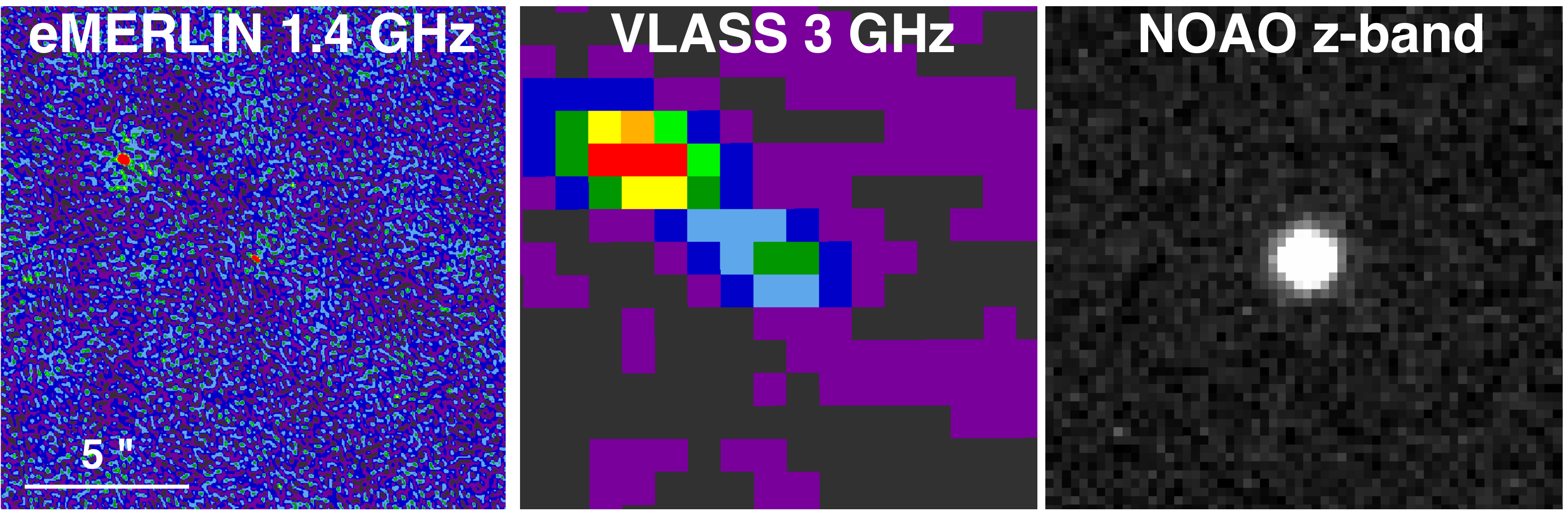}
\caption
{Images of the cQSO 1511+3428 from our e-MERLIN observations (left panel), the VLA Sky Survey Quick Look
(VLASS epoch1.1, middle panel), and
the NOAO Legacy Surveys in the z-band. The source is strongly one-sided; a bright knot is seen to the north-east of the core, but
there is no detectable emission to the south-west. The z-band imaging does not show any optical source at the location of the north-east
knot, so we cannot verify whether it corresponds to a galaxy in the vicinity of the QSO.
}
\label{multi_image}
\end{figure*}

Here we summarise the properties of these 7 rQSOs and cQSOs with clearly extended emission. High-contrast images covering their full extent
are shown in Figures \ref{extended_sources} and \ref{multi_image}, the multi-component fits are shown in Supplementary Figure 2 of the online material, and their parameters summarised in the lower part of Table \ref{fit_summary}. 

\subsubsection{0946+2548}

The presence of a nearby bright radio source affected the imaging of this object, leading to a high level of noise. 
The final self-calibration was performed on this nearby source, and then applied to the target. 

The QSO radio source is composed of two bright resolved knots separated by $0\farcs5$,
along PA$=-17^{\circ}$ which we take to define an axis for the system. The GAIA position of the QSO
lies between the two main knots, coinciding with another faint knot which has a marginal S/N (Figure \ref{images}).

The brighter north-western knot shows an extension perpendicular to the axis.
There is a hint of diffuse patchy emission oriented along the axis that extends beyond the two knots, 
which may indicate partially-recovered extended emission. We modelled the source using four components:
two for the north-western knot, one for the south-eastern knot, and a large elongated component to capture
the diffuse emission around the knots.

The elongated component dominates the total flux, but it may be influenced by residual {\it uv} noise
in the image. This may explain the $\approx \times2.5$ greater flux that we measure in this source compared to its 
flux from the FIRST survey (Section \ref{fluxcomps})

\subsubsection{1007+2853}

The source consists of two compact components separated by $0\farcs34$ along PA$=0^{\circ}$, which
we modelled using two gaussian components. The QSO position lies between the two components, suggesting
that they are opposite lobes of a bipolar jet or outflow.

\subsubsection{1122+3124}

The source consists of a single dominant core component that shows a faint elongation along PA$=205^{\circ}$. A single
component fit leaves a definite visible residual along the direction of the elongation. Therefore, two components are required to
 fit the source adequately.

\subsubsection{1153+5651}

This source exhibits a complex morphology with a dominant central core and a bipolar extension along PA$=171^{\circ}$.
In addition to the bright unresolved core, there is a mildly resolved knot $0\farcs42$ away, from which a faint tendril of emission
extends even further north. Towards the south, faint emission appears to outline the edge of a poorly-defined lobe. 

We model the source using three components, one each for the core, northern knot and extended jet-like emission. The low S/N
of the apparent lobe-like southern feature prevents it from being modelled.

\subsubsection{1535+2434}

The pipeline-calibrated images of this source were strongly affected by two bright sources a few arcminutes away. 
A first round of wide-field self-calibration was used to model the target as well as these bright nearby contaminants. After strong data averaging
to smear out the effects of the other sources, a final round of self-calibration was performed on the target to obtain the final
image.

The QSO radio source is composed of two bright knots separated by $0\farcs27$
along PA$=115^{\circ}$. The GAIA position of the QSO is close to the centre of the western knot, which is also extended beyond the
size of the beam. 

\subsubsection{1410+2217}

This is one of two cQSOs with detectable extended emission. The source structure is dominated by a bright unresolved core.
South of the core, $2\farcs25$ away along PA$=173^{\circ}$, a faint arc-like feature marks the outer edge of a barely-defined
lobe. We modelled the source as a core component and two faint knots on the edge of this tentative lobe, which 
we manually adjusted and tied during the fit. 

\subsubsection{1511+3428} \label{specialcqso}

The QSO was selected for visual compactness in its FIRST imaging, but our e-MERLIN image clearly resolves it into two compact
sources, of which the fainter source corresponds to the optical location of the QSO (Figure \ref{multi_image}). 
The two knots are separated by $5\farcs06$
along PA$=53^{\circ}$. The core is unresolved, but the north-eastern knot has a FWHM that is 25\% larger than the restoring beam,
indicating a moderately resolved structure.

The brighter north-eastern knot could be one hotspot of an asymmetric FRII-like radio galaxy. Neither FIRST or VLASS images, or a natural-weighted large-scale image of the e-MERLIN data, show any evidence of a south-eastern counterpart to the north-eastern knot. Therefore,
if this is a bipolar radio galaxy, the two lobes would need to have a high contrast of greater than $\approx 70$. The VLASS Quick Look image
shown in Figure \ref{multi_image} suggests a hint of emission between the core and knot, which lends credence to the radio galaxy interpretation.

Moderately deep optical imaging of the region around the QSO from the NOAO Legacy surveys does not show any optical counterpart at
the location of the north-eastern knot, which also supports its hotspot nature. On the other hand, at the depth of the 
optical image ($z=23$ AB mag), the corresponding limit on the equivalent rest-frame U-band luminosity of any galaxy at the redshift
of the QSO ($z=1.439$) yields a stellar mass of $\approx 7\times10^{10}$ \msun, taking a solar-metallicity simple stellar population with an age 
$\approx 1.5$ Gyr, a reasonable model for a radio-loud galaxy host in the environment of the QSO. The limit is only
weakly dependent on the age or metallicity of the assumed stellar population. Clearly, a rather massive galaxy in the vicinity of the
QSO could remain undetected with the available optical imaging, if it was quiescent or dusty. Therefore, the absence of a visible
optical counterpart to the north-eastern knot should not be taken as strong evidence that this object is a large asymmetric 
radio galaxy centred on the QSO.

\subsection{FIRST vs. e-MERLIN radio fluxes} \label{fluxcomps}

\begin{figure}
\centering 
\includegraphics[width=\columnwidth]{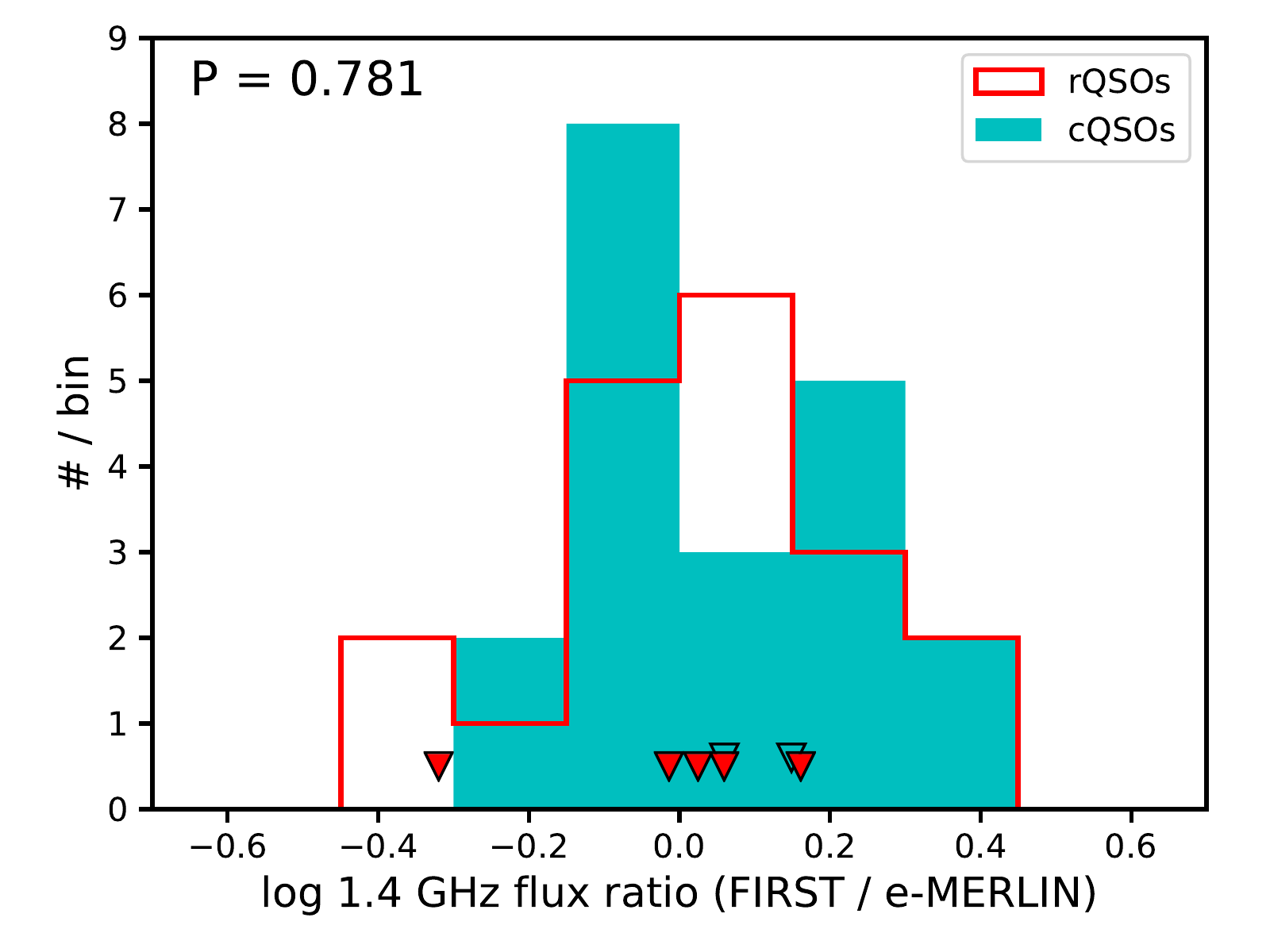}
\caption
{The ratio of the total measured e-MERLIN flux to the FIRST flux for the radio QSOs from our study;
rQSOs (red open histogram) and cQSOs (cyan closed histogram)
are plotted separately. The number in the upper left is the probability that the flux ratios of the
cQSOs and rQSOs are drawn from the same parent distribution as determined by a K-S test. The
downward triangles mark the flux ratios for the five rQSOs (red points) and two cQSOs (cyan points) 
that have extended radio emission in their e-MERLIN images. The cQSO points have been slightly offset
vertically in the Figure only to improve their visibility.
}
\label{fluxratios}
\end{figure}

The e-MERLIN interferometer has an L-band maximum angular scale of $\approx2$": our images are not sensitive to 
radio structures that are larger than these scales. In contrast, the resolution of FIRST is at least 5" across.
Therefore, our radio QSOs could potentially have extended radio emission that is invisible to e-MERLIN,
yet remains unresolved with FIRST.

We can search for evidence of this by comparing the L-band radio fluxes from FIRST to the fluxes we have measured
from the e-MERLIN images through our Gaussian component analysis (Section \ref{gaussfits}). A significant and systematic excess
of FIRST fluxes over e-MERLIN fluxes would suggest radio emission that is resolved out by e-MERLIN.

In Figure \ref{fluxratios}, we plot the FIRST-to-e-MERLIN flux ratio distributions of the rQSOs and cQSOs separately. 
The median ratio for rQSOs is 1.06, compared to 1.02 for cQSOs; the difference is not significant given the substantial
scatter of the observed flux ratios (standard deviations of $\approx 0.2$ dex). 
This is verified through a two-sided K-S test, which indicates
that the two distributions have a $80$\% chance of being drawn from a common parent sample.

The observed scatter of the flux ratios is larger than the estimated flux scale uncertainty of our observations ($<10$\%; Section \ref{obsred}).
Therefore, we attribute these flux differences primarily to radio source variability in these compact QSOs over the $\approx 8$ years
that separate the epochs of the FIRST and e-MERLIN observations. Since the flux ratios of the
rQSOs and cQSOs do not differ significantly, we conclude that neither subsample shows any conclusive
evidence for the widespread presence of bright radio structure extending beyond a few arcsec ($\gtrsim 15$ kpc at the redshifts
of our QSOs).

Having said this, the seven sources with visual extended structure (downward pointing triangular points in Figure \ref{fluxratios}) 
have flux ratios that are generally shifted towards positive values, which may result from low-level extended emission 
from radio lobes that has been resolved out. This emission is unlikely to dominate the radio flux in these sources. The rQSO
with a negative ratio is 0946+2548, where the e-MERLIN flux estimate comes primarily from the large low surface-brightness
component which is poorly constrained, and possibly overestimated. While we can trust the accuracy of the e-MERLIN 
flux estimates for the core-dominated unresolved sources, the resolved sources could have larger errors. As the FIRST
fluxes will be reliable for all our sources, we only use these in the following analyses of radio properties, keeping in mind
that variability sets a systematic scatter of a factor of 2 on \lrad.

\subsection{Radio spectral indices} \label{spectral_indices}

\begin{figure}
\centering 
\includegraphics[width=\columnwidth]{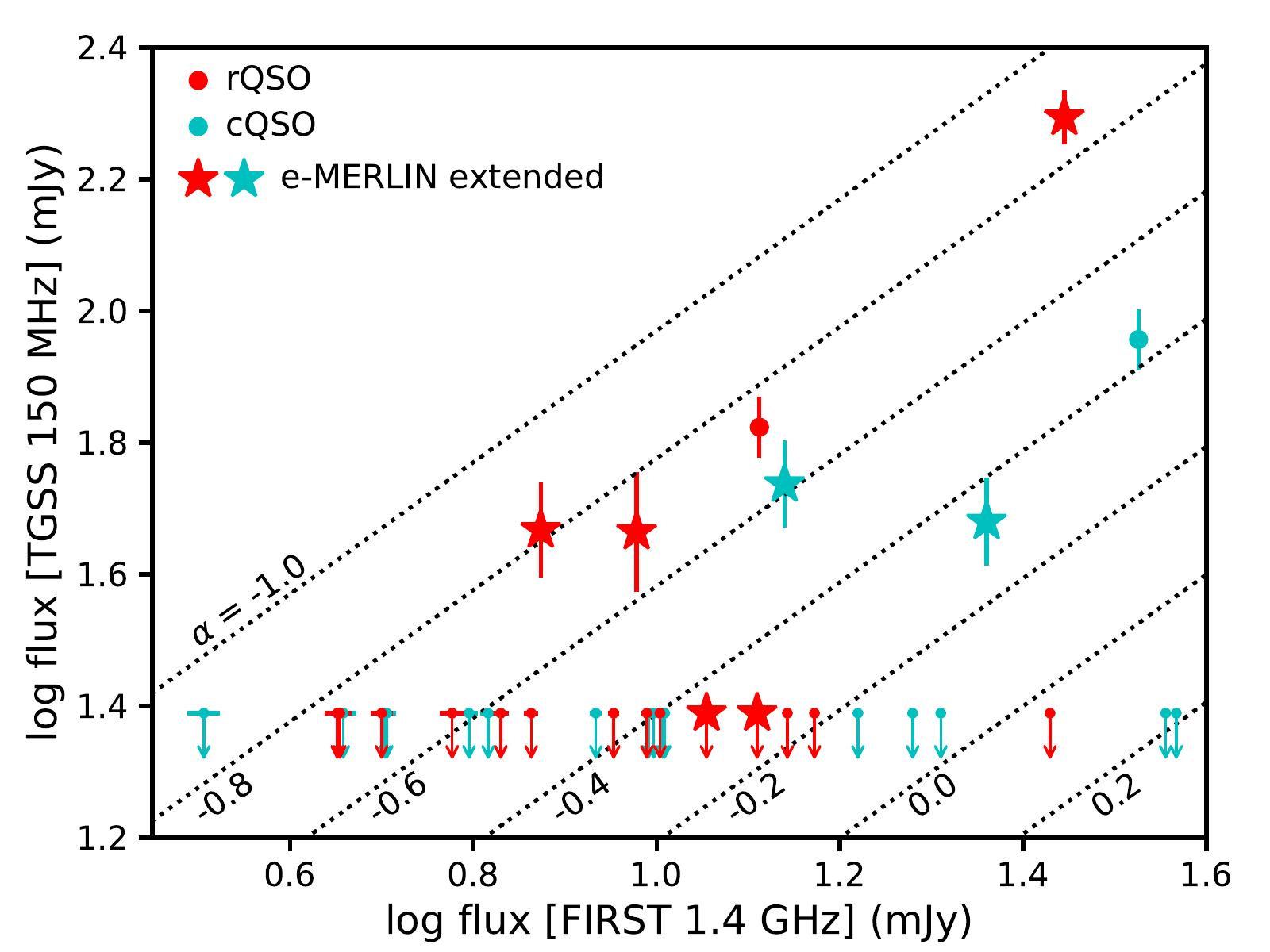}
\caption
{1.4 GHz flux densities from FIRST plotted against 150 MHz flux densities from TGSS for the QSOs in our sample. The diagonal dotted lines
mark tracks of constant radio spectral index between the two frequencies, ranging from an inverted spectrum ($\alpha=0.2$) in the
lower right to a steep spectrum ($\alpha=-1$) towards the middle left. QSOs that are extended in the e-MERLIN
images are shown with larger star-shaped points. The upper limits are plotted at 24.5 mJy, the median detection limit of the TGSS survey.
}
\label{alphaplot}
\end{figure}

Figure \ref{alphaplot} plots the 1.4 GHz and 150 MHz flux densities of the 39 QSOs with coverage in the TGSS ADR (Section \ref{tgss_radio}).
The dotted diagonal lines delineate tracks of constant radio spectral index $\alpha$, where the flux density at a radio 
frequency $\nu$ is given by S$_{\nu} \propto \nu^{\alpha}$. 

The spectral indices of the 8 TGSS-detected QSOs span values from -0.4 to -0.8, fairly typical for AGN-powered radio sources. 
For the rest, we show typical 7$\sigma$ ($=24.5$ mJy) upper limits on the 150 MHz flux densities. These span a large range of possible spectral
indices, and some of the QSOs that are bright in FIRST are expected to have flat or inverted spectra, characteristic of 
very compact core-dominated radio sources. Conversely, the 150 MHz fluxes of the TGSS-detected sources could arise
from very extended steep-spectrum lobes that do not appear in either FIRST or NVSS images. Future LOFAR public
survey data will help discriminate between compact and extended low frequency radio components.

6 of the TGSS detections also show extended emission in their e-MERLIN images, which is consistent with their nature as more
evolved radio sources (Section \ref{jetcontext}). The two rQSOs with extended structure that remain undetected in TGSS (1122+3124
and 1535+2434) are also the most core-dominated, and their limits suggest a typical flat-spectrum core.

\subsection{The mid-infrared -- radio plane} \label{mir_radio}

\begin{figure*}
\centering 
\includegraphics[width=\textwidth]{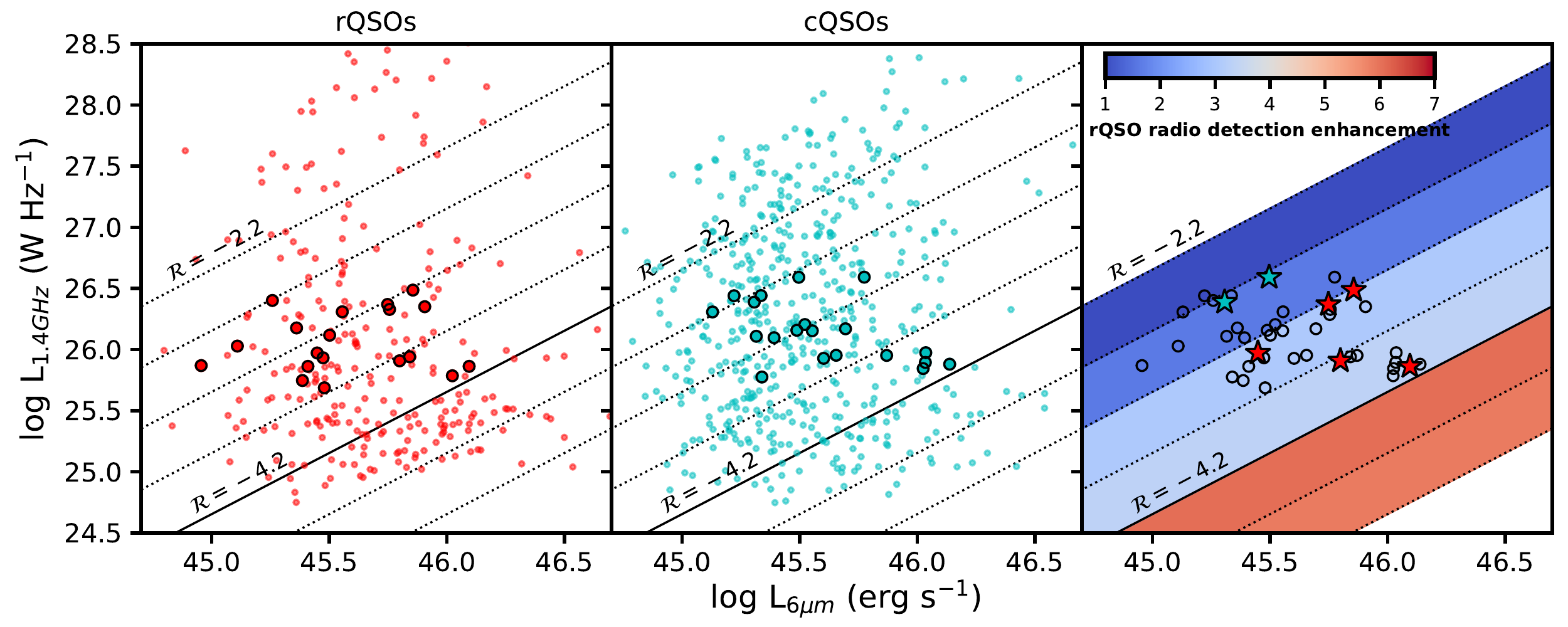}
\caption
{\lsix\ vs. \lrad\ for rQSOs (left panel) and cQSOs (middle panel) from the parent sample (small coloured points) in the
redshift range $1.0 < z < 1.55$ used in this study. The e-MERLIN targets are shown as large coloured points in the respective
panels. Diagonal lines in all three panels
show lines of constant `radio-loudness' $\mathcal{R}$, as defined in Section \ref{mir_radio}. The solid diagonal line marks the
approximate transition between `radio-loud' and `radio-quiet' sources. In the right panel, we plot
the relative enhancement in the radio detection rate of rQSOs vs. cQSOs in the indicated bins of $\mathcal{R}$, as measured
from the very same populations of QSOs plotted in the left and middle panels. The detection rate enhancement increases
towards low values of $\mathcal{R}$ \citep[see also][]{klindt19}. The points plotted in the right panel are the e-MERLIN
targets of this study, and the sources with visually extended emission are distinguished as red star points (five rQSOs) and
cyan star points (two cQSOs).
}
\label{lsix_lradio}
\end{figure*}

In Figure \ref{lsix_lradio}, we plot the MIR luminosity (\lsix) against the FIRST-based rest-frame 1.4 GHz luminosity (\lrad) of colour-selected
QSOs from the parent sample in our working redshift interval. rQSOs and cQSOs are plotted in the left and middle panels 
respectively. As noted in our earlier works, there is no clear relationship between MIR and radio luminosities, with both spanning
orders of magnitude within the parent sample \citep{klindt19, rosario20, fawcett20}. However, there is a clear
difference between cQSOs and rQSOs in the distribution of radio luminosities in this plane. This is evident if we compare
the fraction of cQSOs and rQSOs across the \lrad--\lsix\ ratio, a quantity, when converted into $\log_{10}$ units, 
that we define as the `radio-loudness' $\mathcal{R}$. We refer the reader to \citet{klindt19} and \citet{rosario20}
for a detailed discussion of the $\mathcal{R}$ parameter in relation to traditional radio-loudness measures, but highlight
here that the value of $\mathcal{R}=-4.2$ corresponds to the typical radio-loud/-quiet divide,
such that more radio-loud systems have a higher value of $\mathcal{R}$.

In all three panels of Figure \ref{lsix_lradio}, we plot diagonal lines which mark constant values of $\mathcal{R}$. The solid diagonal line corresponds to the characteristic threshold value of $\mathcal{R}=-4.2$. Earlier work from our team \citep{klindt19, rosario20}
has shown that radio-intermediate rQSOs have a significantly higher incidence within the QSO population, i.e., the enhancement in the radio
detection rate of rQSOs over that of cQSOs is highest for systems that have an $\mathcal{R}$ value around the solid line
in Figure \ref{lsix_lradio}. We can demonstrate this visually by splitting the rQSOs and cQSOs in the left and middle
panels into bins of $\mathcal{R}$ and plotting their relative numbers, normalised by the size of the respective
parent sub-samples to account for the larger number of cQSOs drawn from the parent sample.   
This is shown in the right panel of Figure \ref{lsix_lradio}. Among the most radio-loud QSOs, towards the top left of the panel, 
the relative incidence of rQSOs
and cQSOs are similar, but moving towards the bottom right and decreasing in $\mathcal{R}$, rQSOs become significantly
more numerous with respect to cQSOs until the radio detection limit of FIRST cuts in. Around the radio-loud/-quiet boundary,
there are $\approx 5$ times more rQSOs detected in the radio than cQSOs if one considers equally populated subsets of
the parent sample.

We can now examine the trends shown by our e-MERLIN targets on the MIR-radio plane within the context of the co-eval 
parent QSO population. The targets are shown as large circular points in their respective panels on Figure \ref{lsix_lradio}. 
The narrow range in radio luminosity used to select our targets ($25.5 < \log L_{\rm 1.4} < 26.5$ W Hz$^{-1}$) is evident, as 
well as the fact that they are matched in \lsix. 

The targets are all plotted together in the right panel of Figure \ref{lsix_lradio}, 
where we have additionally highlighted those sources with extended emission
as large coloured star points. Interestingly, the sources seem to cluster differently
on this plane depending on their classification. The two extended cQSOs lie to the upper left of the range shown by our targets, 
in the domain of $\mathcal{R}$ that mark them as being quite radio-loud. 
On the other hand, the five rQSOs with extended emission lie to the right and bottom right, increasingly
close to the radio-intermediate regime where rQSOs are most over-represented in the overall QSO population. 

\subsection{Physical sizes} \label{sizeR}

\begin{figure}
\centering 
\includegraphics[width=\columnwidth]{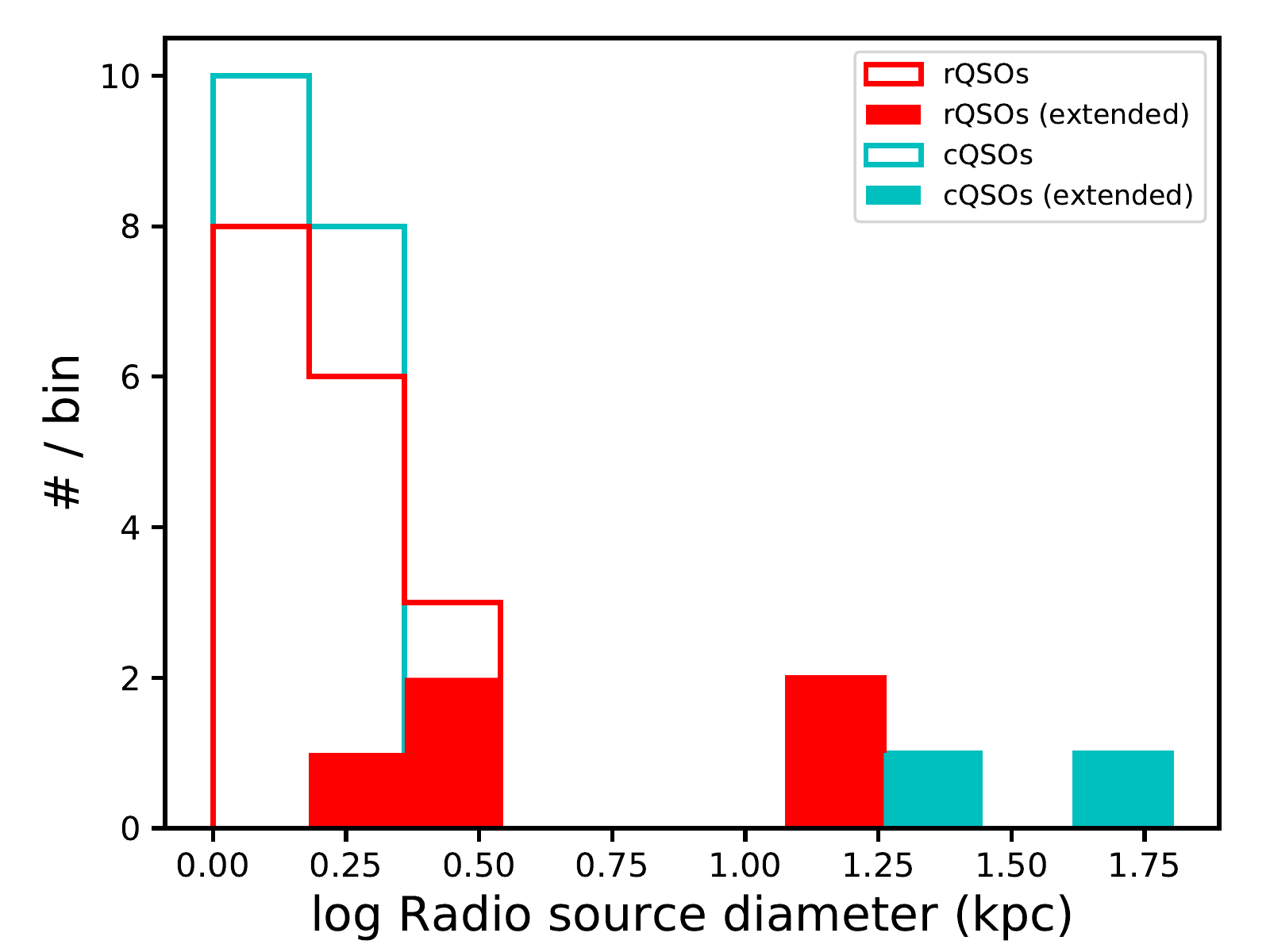}
\caption
{Maximum projected radio source diameters of the QSOs in our sample, measured from gaussian fits to the e-MERLIN images
and converted to physical units. Resolved sources are shown using filled histograms and the limits on the 
core sizes of the unresolved sources are shown using open histograms.
}
\label{sizedists}
\end{figure}

Even a cursory examination of Figure \ref{extended_sources} reveals qualitative differences in the appearance
of the visually extended rQSOs and cQSOs. All five rQSOs show complex, knotty structure, often on small scales,
while the two cQSOs display single unresolved cores and larger hotspots or lobes. This can be quantitatively assessed through a
comparison of the physical sizes of the rQSOs and cQSOs.

We define the maximum size of our sources as either the largest separation between any two Gaussian sub-components 
or the major-axis width of the largest single component, whichever is larger. In the case of sources that only have a single
core component, its major-axis width is adopted as a limit on the source size. We transcribe the measured angular sizes to
physical sizes using the angular diameter distances for the redshifts of our targets. In Figure \ref{sizedists}, we show histograms of the
physical sizes of the ensemble of cQSOs and rQSOs, and use solid histograms to show the sizes of the extended subset.

The two visually extended cQSOs are the largest systems in our sample, both with sizes $\gtrsim 20$ kpc. Two of the extended
rQSOs have similar sizes:  0946+2548 and 1153+5651. The three remaining extended rQSOs are only a few kpc in size, well within
the scale of their host galaxies.

While these differences are interesting, it is important to highlight that the majority of the colour-selected QSOs that we have 
imaged with e-MERLINremain unresolved at scales of $\lesssim 2$ kpc. We interpret this and other results from this work in the 
following discussion.

\section{Discussion} \label{discussion}

Our studies of the SDSS QSO population \citep{klindt19, rosario20, fawcett20}
have conclusively demonstrated that red QSOs (rQSOs) harbour a significantly larger
fraction of compact radio sources of moderate radio-loudness in comparison to normal blue QSOs (cQSOs). Using e-MERLIN,
we have obtained 1.4 GHz (L-band) images of sets of MIR luminosity- and redshift-matched rQSOs and cQSOs in a small redshift
interval ($1.0 < z < 1.55$) with compact morphologies in the FIRST survey.
The new e-MERLIN images achieve angular resolutions of $\approx 0\farcs2$, an improvement of more than an order of magnitude
over FIRST,
which allows a detailed search for resolved radio structure among the very subpopulation of radio quasars in 
which rQSOs and cQSOs differ the most.


We undertake Gaussian single- or multi-component decomposition of the radio images of our targets, and apply a battery of tests
to search for extended structure: visual assessment, core/beam comparisons, and the examination
of positional offsets of the radio core with respect to the optical GAIA positions of the QSOs. We also search for differences in the amount
of resolved 1.4 GHz emission by comparing FIRST and e-MERLIN fluxes. Our analysis only identifies 7 visually extended sources (5 rQSOs
and 2 cQSOs) from the final colour-selected subsets; the remaining 32 QSOs are unresolved with e-MERLIN.


In the rest of this discussion, we collate
the evidence for morphological differences in the radio structures of red and normal QSOs, and present a plausible scenario
that connects the dust that reddens the rQSOs and their radio properties. We also comment broadly on the nature of
compact radio sources in QSOs near cosmic noon, as revealed by this e-MERLIN study. 

\subsection{Kiloparsec-scale radio structures are more common in red QSOs} \label{size_stats}

In terms of the radio structures of the ensemble of QSOs from our study, the clearest result is that the majority of our targets 
are unresolved and therefore exhibit pure core radio morphologies. 

32/40 ($82^{+5}_{-8}$\%) of the QSOs remain unresolved at the physical resolution of the e-MERLIN images ($1.2$--$2.5$ kpc).
Comparisons of the e-MERLIN and FIRST 1.4 GHz fluxes of our targets (Section \ref{fluxcomps}) do not indicate the presence 
of much emission on scales of 10s of kpc that may be missed by e-MERLIN. 

Of the individual subsamples, 5/19 ($26^{+12}_{-7}$\%) of the rQSOs and 2/20 ($10^{+11}_{-3}$\%) of the cQSOs are clearly extended.
As it stands, our results imply a $\approx 2\sigma$ difference in the incidence of extended structure between the two populations,
which we consider to be statistically significant. 

Since only a small fraction of the cQSOs show extended structure, our interpretation of the cQSO 1511+3428 is important.
For the purposes of baseline statistics, we treat it as a single multi-component radio source with a nuclear core coincident with the QSO. 
However, based on
the discussion from Section \ref{specialcqso}, it is possible that the offset hotspot seen in the image is an independent radio source
associated with a neighbouring galaxy. If this were indeed the case, the fraction of cQSOs with extended structure drops to $5^{+9}_{-2}$\%,
differing from rQSOs at the $\approx 3\sigma$ level.

At face value, the low incidence of extended radio structures on scales of a few arcseconds among the cQSOs is
unexpected. There is no {\it a priori} reason that so many QSOs, selected only to be unresolved at $> 5$'', should not
display detectable radio structures between FIRST and e-MERLIN resolutions. For an example of a contrasting result, 
an early MERLIN survey of Jy-level radio quasars \citep{reid95} only finds one source that remains unresolved at 
sub-arcsecond scales, out of a subset of 19 at the same range of redshifts as our e-MERLIN targets. 

However, we are unable to identify work in the current literature that offers an unbiased perspective on the high-resolution 
radio morphologies of QSOs of similar radio luminosities as shown by our targets. The study of \citet{reid95} mentioned above, 
by virtue of its bright radio flux cut, images quasars that are at least two orders of magnitude more luminous than our sample. A high
incidence of bright large-scale lobes is perhaps not surprising in these very powerful systems.
Several studies have targeted interesting subsamples, such as those with known large radio
jets or steep radio spectra \citep[e.g.][]{akujor91, lonsdale93, mullin08} but these have extended radio structure by construction.
There have been some high resolution studies of low-redshift and/or radio-quiet QSOs  \citep[e.g.,][]{kukula98, jarvis19}, which
typically find extended structures in 50-90\% of their sample. However, the targets in these studies are often pre-selected
by properties that may be connected to extended radio jets, such as high-velocity outflows in narrow forbidden lines like 
[O III]$\lambda 5007$ \citep[e.g.,][]{jarvis19}. 

In this sense, our observations offer a valuable reference for the incidence of extended cm-wave radio structures in typical 
QSOs hosting low-power radio sources at $z\sim1$.

\subsection{The incidence of radio sources of different sizes among red and normal QSOs} \label{size_incidence}

\begin{figure}
\centering 
\includegraphics[width=\columnwidth]{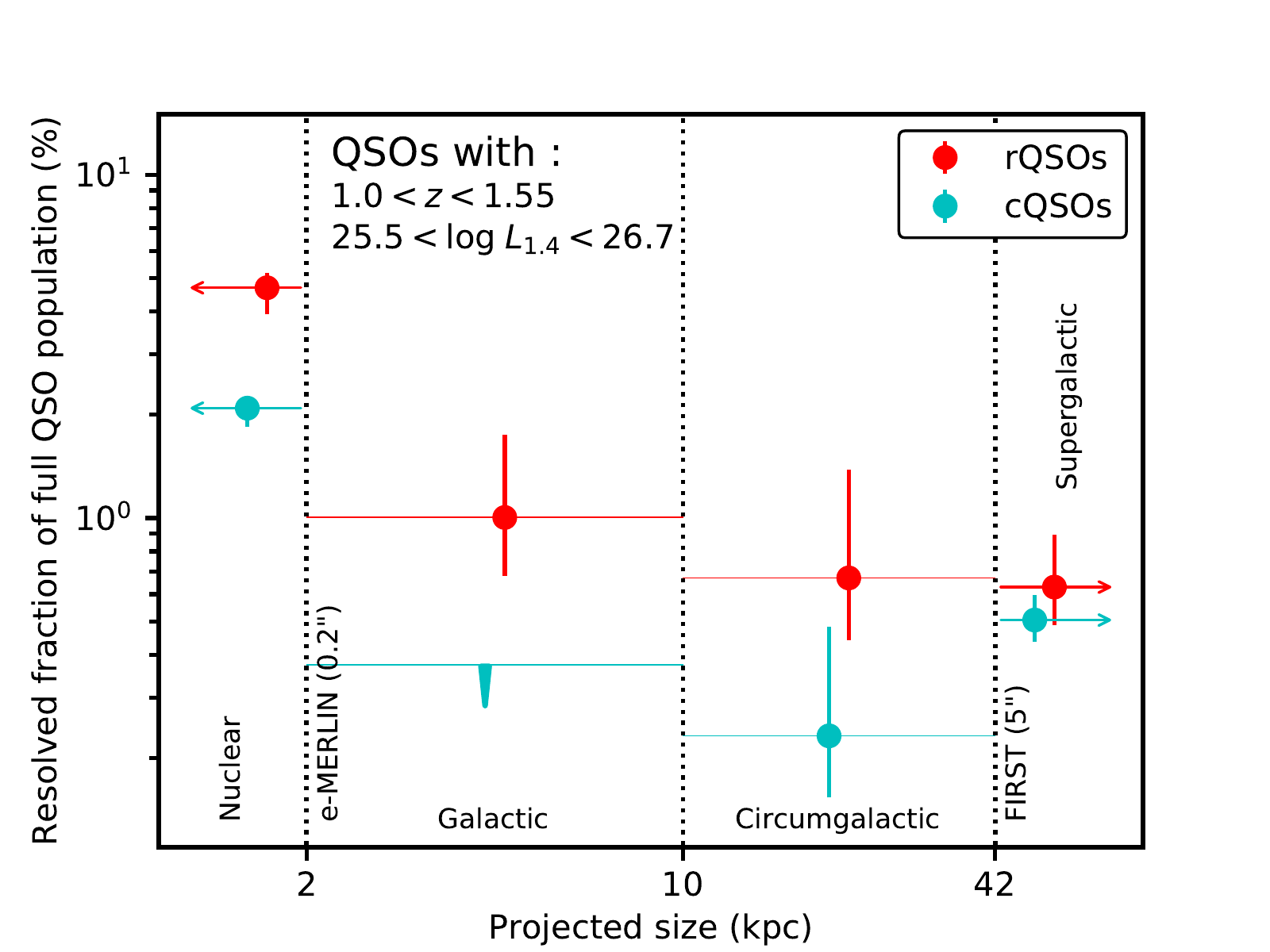}
\caption
{The incidence of radio sources in the \lrad\ range of our targets 
with different maximum projected sizes, expressed as a fraction of the respective colour-selected
QSO subpopulations from the SDSS DR7 with comparable redshifts.
Fractions for the rQSOs (red points) and cQSOs (blue points) are shown separately.
See the associated text in Section \ref{size_incidence} for details.
}
\label{morphdists}
\end{figure}

Our e-MERLIN targets were selected as compact or unresolved sources based on images from the FIRST survey.
The 5'' angular resolution of FIRST translates to projected sizes of $41$--$44$ kpc at the redshifts of our targets, much
larger than the typical sizes of even the massive galaxies that are the likely hosts of these QSOs.
As noted above, it is surprising to find that only a few of our QSOs show resolved radio structure in the e-MERLIN images, despite
our ability to resolve such structure down to $\approx 2$ kpc and with surface-brightness contrasts of $10$--$100$s. 
This implies that radio QSOs show a dearth of emission on scales
of several kpc, approximately in the size range of their host galaxies \citep[e.g.,][]{vdwel14}. 

We investigate this notion further by examining the incidence of radio sources of different (projected) sizes
among QSOs in the redshift and radio luminosity ranges of our e-MERLIN targets ($1.0 < z< 1.55$; $10^{25.5} \leq$\lrad$\leq 10^{26.7}$ \Whz). 
We consider four characteristic physical scales: $<2$ kpc (`nuclear'), 
2--10 kpc (`galactic')\footnote{The choice of a threshold of 10 kpc to separate galactic and circumgalactic scales is 
based on the optical sizes of massive galaxies at $z\sim1.3$, 
the majority of which are found to be smaller than 10 kpc, with a median effective radius of $\approx 5$ kpc \citep[e.g.,][]{vdwel14}.}, 
10--42 kpc (`circumgalactic'), and $> 42$ kpc (`supergalactic')\footnote{The choice of 42 kpc as a threshold between the 
circumgalactic and supergalactic scales is set purely by the resolution limit of FIRST projected to our redshifts of interest.}.

Our approach is to combine statistics from the FIRST survey and our e-MERLIN programme to evaluate, or place limits on,
the fraction of colour-selected QSOs that have radio source sizes within the four ranges defined above. 
An important detail is that these fractions are evaluated with respect to the entire parent population of co-eval colour-selected
SDSS QSOs, not just against the radio-detected subsets. This approach normalises for the key differences in the FIRST
radio properties between rQSOs and cQSOs, {\it viz.} the significantly higher incidence of compact radio sources among rQSOs.

As an illustration, consider the fraction of cQSOs with supergalactic radio sizes. To arrive at this number, we 
first count cQSOs from the parent sample satisfying our redshift and \lrad\ cuts that are also resolved by FIRST.
We then divide this total by the number of all cQSOs in that redshift range irrespective of radio properties. For the three smaller
size bins (nuclear, galactic, circumgalactic), we bring in constraints from e-MERLIN, which, though limited in statistical size,
push well below the FIRST resolution limit. In these cases, we multiply the fraction of cQSOs
that are compact in FIRST by the fraction of e-MERLIN targets that have maximum measured sizes at the respective physical scales, 
propagating uncertainties using binomial statistics \citep{cameron11}.
Similar calculations are performed in each bin separately for rQSOs and cQSOs, and the results are shown in Figure \ref{morphdists}.

We first consider the trend shown among cQSOs, which should be representative of normal QSOs of similar redshift and accretion luminosity. Based on the statistics of e-MERLIN targets that remain unresolved, 
$\approx 2$\% of cQSOs contain nuclear low-power core-dominated radio sources. 
We do not find any cQSOs that have measured e-MERLIN sizes in the 2--10 kpc range, which sets a 2$\sigma$ upper limit of $<0.4$\% 
on the fraction on galactic scales. The 2 e-MERLIN cQSOs that are resolved both have sizes between 10 and 42 kpc yielding a
circumgalactic fraction of 0.1--0.5\%. Using the better statistics for the sources resolved by FIRST, the supergalactic fraction settles to 0.5\%,
comprising of radio galaxies of intermediate radio-loudness with traditional FR I/II classifications, 
as well as more complex extended morphologies \citep{klindt19}.

The trend among the rQSOs reveals important differences when contrasted to that of the cQSOs.
\citet{klindt19} demonstrated that the incidence of FIRST-resolved radio sources
among rQSOs is indistinguishable from cQSOs across a broad range of redshifts. 
Among the QSOs within our working redshift range, the supergalactic fractions for 
both classes are identical within the statistical errors, confirming this result. However,
proceeding towards progressively smaller scales, the fraction of rQSOs rises steadily and deviates strongly from the trends shown
by cQSOs. At galactic scales, the 3 rQSOs that are extended in our e-MERLIN images determine an estimated fraction of $\approx 1$\%, 
more than $\times2$ higher than the conservative upper limit placed on the cQSO fraction. In the nuclear bin, the rQSO incidence is 
$\approx 4.5$\%, again more than $\times2$  the rate found among cQSOs. The significant excess of unresolved radio sources
seen among rQSOs from lower resolution surveys \citep{klindt19, rosario20, fawcett20} can be attributed to a truly compact
population that lies mostly within the inner few kpc of their host galaxies. This indicates that the likely location of the dust that reddens
red QSOs is in the central environment of their AGN. 

%
What can we deduce about the connection between the radio emission and the rQSO phenomenon from these differences in
the fractional radio size distributions? The clearest excess among rQSOs is found among 
resolved and unresolved radio sources at scales $< 10$ kpc. From this, we can conclude that the differences cannot
be purely associated with the synchrotron population in the immediate vicinity (inner few pc) of the accretion disc, such as a jet base or corona.
At least some of these differences must be driven by a radio-emitting structure that extends beyond the scope of the central engine to explain
the differences we find on galactic scales.

\subsection{What is responsible for the excess radio emission in red QSOs?}

Bringing together various threads from the literature and our own analyses, we explore three potential explanations 
for the excess radio emission in rQSOs:
star-formation differences, a higher incidence
of small-scale low-power classical radio jets, or widespread nuclear dusty winds.

\subsubsection{Star-Formation} \label{star_formation}

In a popular evolutionary model connecting red and normal QSOs, rQSOs are caught in a special stage 
during, or shortly after, a powerful merger-sustained starburst event, 
when an explosive AGN feedback episode is clearing out the gas-rich centre of the host galaxy \citep[e.g.,][]{glikman12}. Strong differences in
the star-formation rate (SFR), particularly within the central kpcs, are a natural prediction of this model. Since supernova remnants associated
with star-formation can produce synchrotron emission at GHz frequencies, could the root of the differences we see 
in our e-MERLIN images stem from fundamental differences in the SFR and the spatial distribution of star-forming regions in rQSOs?

The compact radio source components in our e-MERLIN images account for almost all of the integrated radio emission in our targets.
Even the faintest components that we measure from our images have L-band flux densities $> 0.3$ mJy (Table \ref{fit_summary}), 
which, at their redshifts,
imply \lrad\ $>5\times10^{24}$ \Whz, or SFRs greater than a few 1000 \msun\ yr$^{-1}$, taking 
well-used calibrations between SFR and \lrad\ \citep{murphy11, kennicutt12}. Estimates of the SFR from integrated \lrad\ can be
much higher. These derived SFRs are 1-2 orders of magnitude larger than the SFRs of typical QSOs at $z\sim1$ 
\citep{rosario13, stanley17, calistro21}.
Quite simply, the radio luminosities of our sample are too large for star-formation to be an important contributor to the emission we
detect with FIRST or e-MERLIN. Even among co-eval `radio-quiet' QSOs, with \lrad\ fainter than our targets by an order
of magnitude, star-formation is not found to dominate the integrated 1.4 GHz radio emission \citep[e.g.,][]{white17}.
Therefore, we can confidently continue with the notion that the radio differences we see are due to AGN-powered processes.

\subsubsection{Jet-powered radio sources} \label{jetcontext}

\begin{figure}
\centering 
\includegraphics[width=\columnwidth]{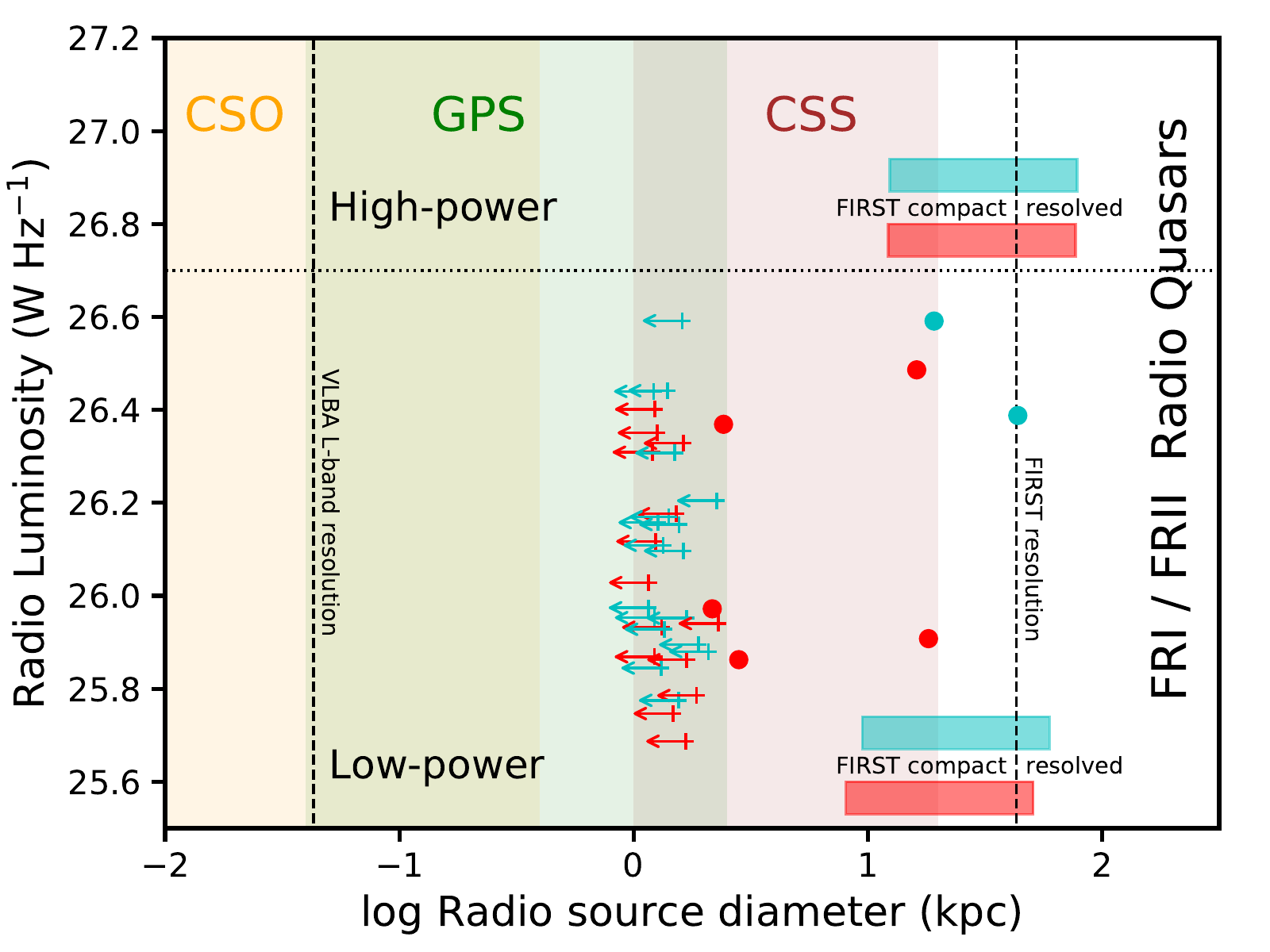}
\caption
{Maximum radio source size vs. 1.4 GHz radio power (\lrad) for our rQSOs (red points) and cQSOs (cyan points).
Filled circles and left-pointing arrows show extended sources and unresolved sources respectively. The dashed vertical
lines mark the typical physical resolution of the FIRST survey and the VLBA at L-band at a characteristic redshift of $z=1.3$.
The dotted horizontal line marks a rough transition between low-power and high-power radio sources.
The coloured bands spanning the FIRST resolution limit graphically show the relative fractions of radio rQSOs and cQSOs 
that are determined to be compact or resolved by FIRST. Note that compact sources are more commons among rQSOs compared
to cQSOs at low radio luminosities, but they are found equally among more luminous systems \citep[e.g.,][]{klindt19}.
We mark using shaded bands the approximate size ranges of the CSS, GPS, and CSO categories of jetted radio sources.
The largest extended radio sources, those that are well-resolved by FIRST, are the classical double-lobed FRI/II systems.
}
\label{size_LR}
\end{figure}

Jet-driven radio galaxies present a plethora of morphologies, and have sizes spanning from 
nuclear scales to some of the largest coherent structures in the Universe \citep{miley80}. Given the long-standing association between 
radio galaxies and QSOs \citep{urry95}, it is reasonable to interpret the sources we have characterised with e-MERLIN
in the context of the population of classical radio galaxies at intermediate redshifts.

The e-MERLIN morphologies of the extended small-scale radio sources in our sample are typically bipolar, but 
they do not all sport the core/jet or core/lobe structure reminiscent of classical double radio galaxies. 
However, it is also well-known that powerful radio sources that are confined within their hosts 
often show quite disturbed morphologies on similar size scales 
\citep[][and references therein]{odea98}.
Therefore, morphological information alone is not sufficient to assess
their nature.

Additional insight and context comes by examining our sources on the space of 1.4 GHz radio luminosity (\lrad) 
vs. the maximum physical size in kpc (Figure \ref{size_LR}).
Similar diagrams have been employed by earlier studies to explore the dynamical evolution of radio sources 
\citep[e.g.][]{odea97,kunert10,an12,jarvis19}. 

As jet-powered radio sources evolve, their primary energy loss mechanisms change 
as they expand and interact with the ambient medium of their host galaxies \citep[][and references therein]{an12}. 
Within the dense, approximately constant density interstellar medium of the central few 100s of pcs of a galaxy, adiabatic losses dominate
over synchrotron losses, leading to a low radiative conversion efficiency and a steep radio spectrum. Such sources are
called Compact Symmetric Objects (CSOs). As they expand beyond the inner galaxy, the confining gas develops a density
gradient decreasing outward and synchrotron losses catch up to adiabatic losses. A prominent spectral
break at cm wavelengths develops and the term Gigahertz-Peaked Spectrum (GPS) is used to describe such sources.
As they burst out of their host galaxies, synchrotron losses take over; jets and lobe expand out into their galactic
environment as the spectral break moves to lower frequencies in the form of Compact Steep Spectrum (CSS) sources
and eventually the classical doubles of FR I/FR II morphologies. 

In Figure \ref{size_LR}, we delineate the ranges of sizes observed in radio galaxies of the various categories outlined above. 
All these categories can span orders of magnitude in \lrad, so we only use their sizes to distinguish them in our diagram,
using various shaded regions which nonetheless have substantial overlap.

The e-MERLIN QSOs that are resolved (circle points) are in the range of sizes shown by CSS sources. This is consistent
with their moderately steep meter-wave spectral indices found in our FIRST-TGSS analysis (Figure \ref{alphaplot}).
The unresolved sources have upper limits to their sizes (left arrow points) that place them 
in the GPS and CSO regime. If their actual sizes are at the order of hundreds of pc, 
we expect these objects to predominantly display synchrotron spectra with strong curvature 
at 100s of MHz to GHz frequencies \citep{odea97}. Unfortunately, from our analysis in Section \ref{spectral_indices},
we cannot clearly ascertain whether the meter-wave spectral indices, most of which are limits, suggest a turn-over or not, but
it is unlikely that the unresolved e-MERLIN sources all show flat or inverted spectra between TGSS and FIRST bands.

\subsubsection{Wind-powered radio sources} \label{windcontext}

Dusty AGN-driven winds  \citep[e.g.,][]{hoenig19} can also offer an explanation for the small-scale radio emission that we observe. 
Recently, \citet{calistro21} presented a comparative multi-wavelength and spectral analysis of rQSOs and cQSOs 
selected using the same approach as our earlier works. They report a correlation between the presence of excess hot
dust emission in the MIR and strong outflow signatures among rQSOs, which is evidence that 
dusty winds are responsible for their red colours. 

Shocks from fast winds in AGN have been invoked in earlier studies to explain enhanced radio
emission in luminous AGN that also show emission-line outflows \citep[e.g.,][]{zakamska14}. The low
radio-loudness of the excess radio sources in rQSOs is consistent with a possible origin in shocks from AGN-driven winds 
that also carry enough dust to mildly redden the nuclear spectra in these objects \citep{rosario20}.

The radio morphologies produced by such winds are still unclear. They are likely to be bipolar,
reflecting the collimation of a wind towards the polar axis of the accretion structure \citep{williamson20}, but
whether they retain their shape over kpcs in the gas-rich environment surrounding the nucleus is subject to debate 
\citep{wagner13}. For e.g., radio-quiet QSOs with strong kpc-scale outflows, systems in which wind shocks are expected to be 
commonplace, appear to show radio structures that look like classical radio galaxies \citep{jarvis19}.

We stress here that it is the {\it excess} compact radio emission found in rQSOs that we seek to understand. 
Nuclear radio sources are seen in our cQSOs as well, and these could very well arise from jets or coronal emission
since cQSOs are not expected to harbour widespread dusty winds. Since nuclear radio sources are about twice as common
among rQSOs as cQSOs (Figure \ref{morphdists}), roughly half of them may arise from standard jet-powered processes. It is
the other half, powered by dusty winds, which accounts for the overall enhancement of radio emission in rQSOs in this scenario.

Very long baseline interferometry (VLBI) has the capacity to resolve the small-scale structure of our sources ($< 100$ pc). 
For e.g., the Very Long Baseline Array (VLBA) has a resolution at 1.4 GHz of 5 mas, probing well into the domain of physical sizes that
separate GPS and CSO sources (Figure \ref{size_LR}). Future multi-frequency studies with this or complementary facilities may allow us to 
explore whether the nuclear radio structures in the majority of our targets could be described as the 
early evolution of bipolar jets. 

Another potential test to distinguish between jet-powered or shock-powered explanations
 is through the use of radio spectral curvature constraints.
Jet-driven double radio sources show strong correlations between their spectral turnover frequencies and their source sizes \citep{odea97},
a consequence of the dynamic changes that such radio sources undergo as they expand within their host galaxies. Radio structures
from wind shocks are not expected to evolve in a similar fashion, and therefore may not lie along such trends. 

With multi-band Giant Metre-Wave Radio Telescope (GMRT) observations which have recently been taken for all 
40 e-MERLIN targets (Principal Investigator: V.~Fawcett),
future work from our team will present a comparative radio spectral and morphological analysis, seeking further
insight into the phenomena that fundamentally distinguish red and normal QSOs.

\section{Conclusions} \label{conclusions}

We present e-MERLIN L-band images of a sample of 19 red QSOs (rQSOs) and 20 normal QSOs (cQSOs) at redshifts of $1.0<z<1.55$.
Our targets are matched in redshift and rest-frame 6 \mics\ luminosity (\lsix), a measure of their nuclear bolometric power, and selected
to have moderate 1.4 GHz luminosities ($10^{25.5} \leq$\lrad$\leq 10^{26.7}$ \Whz) and compact morphologies in the FIRST
survey (angular resolution of 5''; $> 40$ kpc at the QSO redshifts). The e-MERLIN images resolve radio structures down to 
$\approx 0\farcs2$, probing host galaxy scales in these QSOs ($2$--$10$ kpc).

Following a morphological assessment of the e-MERLIN images, both visually and by means of gaussian decomposition, we assessed
the incidence and distribution of radio sources in our targets across a range of projected size scales. Our main conclusions are
as follows:

\begin{itemize}

\item The majority of both cQSOs and rQSOs have nuclear radio cores that remain unresolved with e-MERLIN (Figure \ref{images}
and Section \ref{size_stats}). 

\item We find a statistically significant excess in the incidence of small-scale radio sources (sizes $< 10$ kpc) among rQSOs, while
confirming that larger radio sources are as common in rQSOs as in cQSOs in the redshift and \lrad\ range of 
interest (Section \ref{size_incidence}). 
We report a very low incidence ($<0.4$\%) of sources with galactic-scale sizes ($2$--$10$ kpc) among typical QSOs (cQSOs).

\item Incorporating TGSS constraints, most of the sources with extended e-MERLIN structure also show typical steep radio
spectral indices (Section \ref{spectral_indices}). 

\item Unlike the cQSOs, rQSOs with extended e-MERLIN emission tend to be of intermediate radio-loudness (Section \ref{mir_radio}).

\item We consider various scenarios to explain the excess in kpc-scale radio structures we find among rQSOs. Current constraints
from radio morphologies and spectral indices imply that the differences cannot arise purely from the accretion structure
(the accretion disc and/or corona), but must entail differences in components that partially extend out into the host galaxy. These
could either be classical jet-powered radio sources, or those powered by winds. Since the key feature that defines rQSOs 
is the presence of moderate dust extinction towards their nuclei, we postulate that dusty winds are widespread among
rQSOs, and the shocks that these winds drive into the interstellar medium of their host galaxies are responsible for the
particular radio properties of red QSOs. 

\end{itemize}

\section*{Acknowledgements}
We thank the anonymous referee for input that has improved the material and message of this work.
DJR and DMA acknowledge support from STFC (ST/T000244/1).
JM acknowledges financial support from the State Agency for Research of the Spanish MCIU through the ``Center of Excellence Severo Ochoa'' award to the Instituto de Astrof\'isica de Andaluc\'ia (SEV-2017-0709) and from the grant RTI2018-096228-B-C31 (MICIU/FEDER, EU)
e-MERLIN is a National Facility operated by the University of Manchester at Jodrell Bank Observatory on behalf of STFC, part of UK Research and Innovation.
Funding for the SDSS and SDSS-II has been provided by the Alfred P. Sloan Foundation, the Participating Institutions, the National Science Foundation, the U.S. Department of Energy, the National Aeronautics and Space Administration, the Japanese Monbukagakusho, the Max Planck Society, and the Higher Education Funding Council for England.  
This work has made use of data from the European Space Agency (ESA) mission
{\it Gaia} 
processed by the {\it Gaia}
Data Processing and Analysis Consortium (DPAC,
Funding for the DPAC
has been provided by national institutions, in particular the institutions
participating in the {\it Gaia} Multilateral Agreement.

\section*{Data Availability}

The e-MERLIN data used in this work is available from the corresponding author by reasonable request. All other data
are in the public domain and accessible through the corresponding data archives. All measurements required to replicate
our results are made available in this paper in the form of figures and tables.

\bibliographystyle{mn2e}

\bibliography{rqso_emerlin}

\end{document}